\documentclass[usenatbib]{mn2e}

\usepackage{rotating}
\usepackage{lscape}
\usepackage{subfigure}
\usepackage{graphicx}
\usepackage{longtable}
\usepackage{float}
\usepackage{captcont}
\usepackage{endfloat}
\usepackage{morefloats}
\usepackage{mathrsfs}
\usepackage{alphalph}

\usepackage{multicol}
\usepackage{multirow}
\usepackage{layout}

\defcitealias{noterdaeme10}{N10}
\defcitealias{york12}{Y12}
\defcitealias{straka13}{S13}
\defcitealias{srianand13}{Sr13}

\title[Galaxies on top of QSOs]{
Galactic nebular lines in the fiber spectra of background QSOs: Reaching
a hundred QSO-galaxy pairs with spectroscopic and photometric
measurements}

\author[Lorrie A. Straka]
{Lorrie A. Straka,$^1$ Pasquier Noterdaeme,$^2$  Raghunathan Srianand,$^3$ 
\newauthor
Songkiat Nutalaya,$^4$ Varsha P. Kulkarni,$^5$  Pushpa Khare,$^6$ David Bowen,$^7$
 \newauthor
 Michael Bishof,$^1$ Donald G. York$^{1,8}$ \\
$^1$Department of Astronomy and Astrophysics, University of Chicago, Chicago, IL 60637, USA\\
$^2$Institut d'Astrophysique de Paris, UMR7095, CNRS-UMPC, 98bis bd Arago, 75014 Paris, France \\
$^3$Inter-University Centre for Astronomy and Astrophysics (IUCAA), Ganeshkhind, 411\,007 Pune, India\\
$^4$Department of Astronomy, University of Massachusetts, Amherst, MA 01003, USA\\
$^5$Department of Physics and Astronomy, University of South Carolina, Columbia, SC 29208, USA\\
$^6$CSIR Emeritus Scientist, IUCAA, Ganeshkhind, Pune 411\,007, India \\
$^7$Department of Physics and Astronomy, Princeton University, Princeton, NJ 08544, USA\\
$^8$Enrico Fermi Institute, University of Chicago, Chicaglo, IL 60637
}

\date{}

\pagerange{\pageref{firstpage}--\pageref{lastpage}} \pubyear{}

\def\LaTeX{L\kern-.36em\raise.3ex\hbox{a}\kern-.15em
    T\kern-.1667em\lower.7ex\hbox{E}\kern-.125emX}

\begin{document}

\maketitle

\label{firstpage}

\begin{abstract}
We present photometric and spectroscopic measurements of 53 QSO-galaxy pairs from the Sloan Digital Sky Survey, where nebular emission lines from a $0<z<0.84$ foreground galaxy are detected in the fiber spectra of a background QSO, bringing the overall sample to 103 QSO-galaxy pairs detected in the SDSS.  We here study the nature of these systems. Detected foreground galaxies appear at impact parameters between 0.37 kpc and 12.68 kpc. The presence of oxygen and Balmer emission lines allows us to determine the emission line metallicities for our sample, which are on average super-solar in value. Star formation rates for our sample are in the range $0.01-12$ M$_{\odot}$\,yr$^{-1}$. We utilize photometric redshift fitting techniques to estimate the M$_{\ast}$ values of our galaxies (log M$_{\ast}$ = 7.34 - 11.54), and extrapolate this relationship to those galaxies with no imaging detections. Where available, we measure the absorption features present in the QSO spectrum due to the foreground galaxy and the relationships between their rest equivalent widths. We report an anti-correlation between impact parameter and E(B-V)$_{(g-i)}$, as well as a correlation between galaxy color ($u-r$) and E(B-V)$_{(g-i)}$. We find that our sample is one of late-type, star forming galaxies comparable to field galaxies in a similar redshift range, providing important clues to better understand absorption systems. These galaxies represent a sample of typical galaxies in the local Universe for which abundances, extinction, morphology, and absorption properties may be measured using background QSOs with great potential for follow-up observations.

\end{abstract}


\section{Introduction}\label{section-introduction}

Various techniques have been developed to uncover the links between gas and galaxies over cosmic history, 
approaching the problem from either side. For example, the gas { causes a unique absorption signature to be imprinted in the 
spectra of background sources,} which, at low redshift, can be associated and correlated with the 
properties of foreground galaxies \citep[e.g.][]{bergeron91}. At high-redshift, absorption 
lines have been used to predict properties of the associated galaxies such as column density, absorption metallicity, abundance ratios, and gas kinematics \citep[e.g.][]{ledoux98, wolfe03, srianand05} where direct detection of the associated galaxies remains challenging and limited to a small number of cases \citep[e.g.][]{noterdaeme12, peroux12, fynbo13}. On the other hand, cross-correlating low redshift galaxy and QSO position catalogs can be used to 
subsequently study the properties of the gas \citep[e.g.][]{gupta10}. 

While absorption lines offer a powerful tool for studying the evolution of galaxies over 
a range of epochs, QSOs with known foreground galaxies do not necessarily feature the expected 
absorption signatures \citep[e.g.][]{bowen95, tripp05, chen10}. 
Therefore, it is important to investigate the properties of those 
foreground galaxies known in emission in order to correlate them with the presence or absence of absorption, and with the properties of absorption features, if detected. 
Do these 
systems select the most metal-rich gas \citep[e.g.][]{wild06, nestor08}, and are these 
halo gas detections?  The question yet remains how absorption systems seen in QSO spectra are related to the 
emission characteristics of the underlying galaxies. 

Recently, narrow galactic emission lines present in the fiber spectra of background 
QSOs have been successfully utilized to probe galaxies at relatively small impact parameters. The 
techniques for galactic emission line searches are presented in \cite{borthakur10}, \citet[hereafter N10]{noterdaeme10}, \citet[hereafter Y12]{york12}, and in \citet[hereafter S13]{straka13}. The selection of galaxies via emission lines is independent of its 
absorption characteristics and broad-band fluxes, bringing a valuable, independent view on the relationship 
between galaxies and absorption systems. 

N10 presented a sample of [O\,{\sc iii}]-emitting galaxies on top of QSOs (GOTOQs). Therein, they focus on 
a subset of 17 systems at $z=0.4-0.8$ for which the Mg\,{\sc ii} doublet is covered by the spectra. 
They measured lower limits on the SFR in the range 0.2-20 M$_{\odot}$ yr$^{-1}$ and detected 
strong Mg\,{\sc ii}, Mg\,{\sc i}, and Fe\,{\sc ii} absorption lines in most cases. 
There is also a possible 2$\sigma$ correlation between emission-line metallicity and Mg\,{\sc ii} 
rest equivalent width. By stacking spectra of a large sample of Mg\,{\sc ii}-selected systems, N10 
also detected for the first time a correlation between Mg\,{\sc ii} equivalent widths and the fluxes in the [O\,{\sc ii}] and [O\,{\sc iii}] lines. {  However, the interpretation that it represents a correlation between Mg\,{\sc ii} equivalent width and SFR \citep{menard11} may be too simplistic as differential fiber losses are at play \citep{lopez12}. These are based on a stacked sample of higher redshift absorbers and the geometry of the absorption and emission is not clear \citep{menard12}}.

Y12 dealt with the detection of GOTOQs in SDSS DR5 through a search for strong H$\alpha$ and presented methods 
to deconvolve the QSO from the galaxy in the SDSS images.
Building upon Y12, S13 utilized a multi-emission line search, simultaneously matching up to nine galactic 
emission lines in the optical. This paper showed that GOTOQs are more likely to be 
late-type, star forming, and lightly reddened galaxies. \citet[hereafter Sr13]{srianand13} presented results on one QSO-galaxy pair in 21-cm and optical data. They found that this line of sight has near-solar emission metallicity and dust extinction typical of translucent clouds. They also detect the $\lambda 6284$ diffuse interstellar band (DIB) and strong 21-cm absorption at the redshift of the galaxy.

Here we present 28 additional detections of emission lines from 
foreground galaxies in QSO spectra, obtained by applying the procedures of N10 to H$\alpha$. 
We perform spectroscopic and photometric measurements as in S13 for previously known GOTOQs 
(mostly from N10) for which such an analysis was not yet performed. 

{ Results showing followup observations of such cases of GOTOQs may be found in \cite{borthakur10}, \cite{gupta13} and \cite{srianand13}. All three papers report the results of galactic interstellar absorption from HI 21cm, Ca II H and K, and the Na I D lines.}

Section~\ref{section-sample} details our sample selection and addresses any bias in the sample. Section~\ref{section-analysis} details our data analysis. Section~\ref{section-results} reports our results. Finally, section~\ref{section-conclusion} summarizes our conclusions. Throughout we have assumed the concordance cosmology (H$_{0}=70$ km s$^{-1}$ Mpc$^{-1}$, $\Omega_{m}=0.30$, $\Omega_{\Lambda}=0.70$).

\section{Sample and Bias}\label{section-sample}

\subsection{Sample Selection}

We have systematically searched 102,418 QSO spectra from SDSS DR7 for intervening H$\alpha$ emission 
(excluding $\sim2000$ for poor signal-to-noise) at $0<z<0.35$. The QSO continua for these spectra 
were iteratively fit by applying Savitzky-Golay filtering \citep{savitzky64}, which performs a local polynomial regression 
while preserving relative maxima and minima (such as emission and absorption features). Then the continuum-subtracted QSO spectra were cross-correlated with a template profile of 
H$\alpha$+[N\,{\sc ii}] {  (referred to as ``H$\alpha$'')} generated from the SDSS galaxy template spectra. No signal-to-noise restrictions 
were placed on the H$\alpha$ search. About 350 H$\alpha$ candidates were returned from this search. 
Each candidate was then visually inspected for the presence of other galactic emission lines, such as 
H$\beta$ and [O\,{\sc ii}]$\lambda3727$. This provided us with { 52} GOTOQs (including 17 $z>0.4$ Mg II absorbers from N10), plus the { 50} found with the independent searches from Y12 (H$\alpha$) and S13 (multiple emission). In addition, one system has also been studied in \cite{gupta13}, who present a study of the 21-cm absorption it produces in the spectrum of the background QSO. We include this system here due to its classification as a GOTOQ according to our criteria. This brings our total sample to { 103} GOTOQs. { We present spectroscopic and photometric measurements for the 53 remaining systems (the first 50 being published in Y12 or S13). } 

Throughout the paper, we primarily refer to the total sample of 103 GOTOQs. We will also refer to { three sub-samples. From our total sample, all objects selected by H$\alpha$ will be represented in the figures as plus signs and referred to as the H$\alpha$ sample, all objects selected by [O\,{\sc iii}] will be represented as x's in the figures and referred to as the [O\,{\sc iii}] sample, and those objects selected simultaneously by multiple lines will be represented as dots in the figures and referred to as the multi sample.}  

\begin{figure}
\begin{minipage}{1.0\linewidth}
\begin{center}
\subfigure[]{
\includegraphics[trim = 0mm 0mm 0mm 0mm, clip, width=0.5\linewidth]{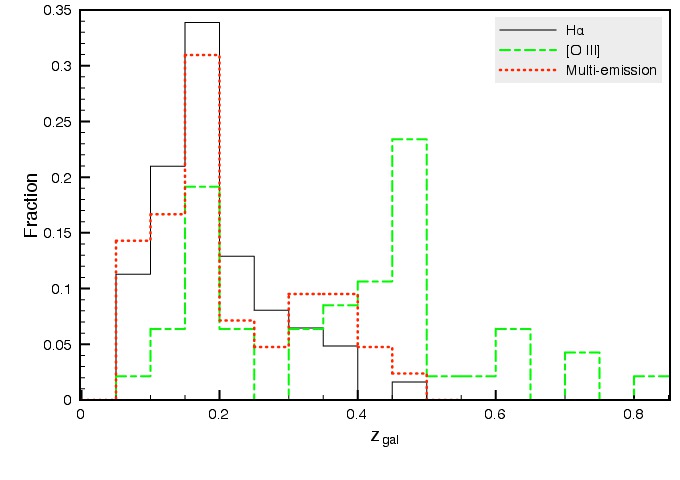}
\label{fig-redshift-dist}
}
\subfigure[]{
\includegraphics[trim = 0mm 0mm 0mm 0mm, clip, width=0.5\linewidth]{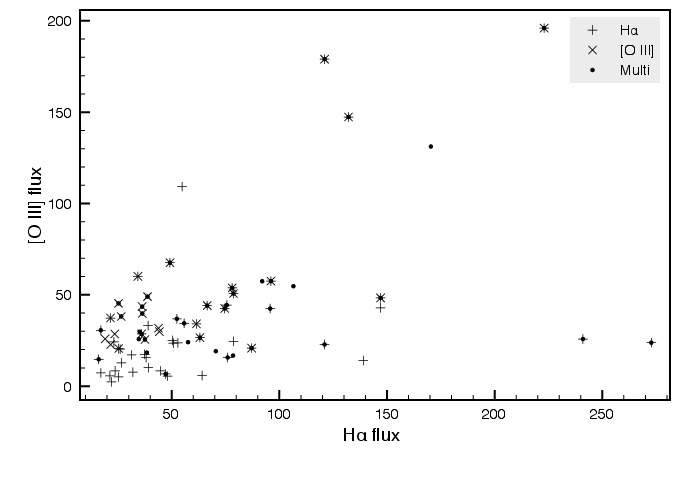}
\label{fig-flux-dist}
}
\end{center}
\caption{a) The redshift distribution for our galaxy sample. Redshifts are determined by galactic emission in multiple lines. We see the same peaked distribution for the three search techniques at $z<0.4$, with extended coverage for the [O III] search, which was able to reach up to $z=0.8$. b) [O III] flux vs. H$\alpha$ flux for or sample. We see no difference in the detection techniques, which return roughly the same flux distribution in both H$\alpha$ and [O III].}
\end{minipage}
\end{figure}

Figure~\ref{fig-redshift-dist}a shows the redshift distribution of the total galaxy sample divided by detection criteria (H$\alpha$, [O III], or multiple-emission) and Figure~\ref{fig-flux-dist}b shows the [O III] flux vs. H$\alpha$ flux, again plotted by detection criteria. We find no discernible difference between the three SDSS GOTOQ samples in terms of parent population, 
indicating the selection methods are roughly equivalent. { The three different search methods used here return slightly different sets of unique systems. However, we find distributions of various properties of these systems are similar. This implies that the three galaxy samples are drawn from the same population despite being selected through different search techniques, as can be seen in the figures.}

{ Table~\ref{tbl-total-sample} lists our sample of 103 GOTOQs: the SDSS plate, fiber, and MJD; the QSO RA and Dec; references where the GOTOQ was first reported and where it were first photometrically treated in publication; and which of the three search functions returned them (allowing us to compare the three selection techniques). Those GOTOQs present here (53 of them) are labeled as this work.  All subsequent tables and figures deal only with those targets treated in this work. Table~\ref{tbl-phot-quasars-III} lists the index number for each of the 53 targets in this work, by which they will be referred to throughout, and the truncated four digit RA and Dec ID for each target, the SDSS r-band PSF magnitude for the QSO, the Petrosian radius of the QSO, the five filter deconvolved magnitudes of the QSO, the reddening estimates E(B-V)$_{(g-i)}$ due to the foreground galaxy, and the number of other QSOALS along the line of sight to each QSO (excepting the galaxy studied in this work).} 

{  We have included online only a machine readable table of all the measurements taken for the total sample. The first row includes the column headers. The columns for the table are as follows: QSO RA, QSO Dec, plate, MJD, fiber, QSO z, galaxy z, luminosity distance,  three selection columns (H$\alpha$, [O III], and multi-line detection marked '1' for each if they are found with these techniques), visual detection, Petrosian radius, PSF r-band magnitude of the QSO, QSO u magnitude, QSO g magnitude, QSO r magnitude, QSO i magnitude, QSO z magnitude, the number of additional intervening QSOALS, $\theta$ (arc seconds), impact parameter (b; kpc), galaxy u magnitude, log L/L$_{\odot}^{u}$, galaxy g magnitude, galaxy r magnitude, log L/L$_{\odot}^{r}$, L$^{\ast}$/L$_{\odot}^{r}$, galaxy i magnitude, log L/L$_{\odot}^{i}$, galaxy z magnitude, (u-r) galaxy color, E(B-V)$_{g-i)}$, H$\alpha$ flux, error, H$\beta$ flux, error, [O II] flux (unresolved), error,  [O III]a flux, error,  [O III]b flux, error, [N II]a flux, error, [N II]b flux, error, [S II]a flux, error,  [S II]b flux, error, R$_{23l}$, R$_{23u}$, O3N2, N2, E(B-V)$_{H\alpha/H\beta}$, SFR$_{H\alpha}$, SFR$_{[O II]}$, SFR$_{H\alpha}^{corrected}$, SFR$_{[O II]}^{corrected}$, log M$_{\ast}$, 1$\sigma$ M$_{\ast}$ low bound, 1$\sigma$ M$_{\ast}$ upper bound,  Ca II K equivalent width (EW), error, Ca II H EW, error,  Na I D2  EW, error,  Na I D1 EW, error,  Mg II $\lambda$2796 EW, error, Mg II $\lambda$ 2804 EW, error, Mg I $\lambda$ 2852 EW, error, Fe II $\lambda$2587 EW, error, and Fe II $\lambda$ 2600 EW.  Dashes in the cells indicate measurements that could not be made due to missing data (redshift range or missing data). Each flux and absorption measurement column have error columns following them.  Fluxes are measured in $10^{-17}$ ergs s$^{-1}$ cm$^{-2}$. Absorption lines are rest equivalent width. Magnitude errors are less than 0.05 mag in each band. Limits are discussed in Section~\ref{section-photometry}. }

\subsection{Bias}

The discovery of this class of objects was serendipitous \citep{borthakur10, noterdaeme10, york12}. The followup to these discoveries  (N10, Y12, and S13) was aimed at finding interesting galaxies for which further studies (21cm, QSOALS) would provide insights into the nature of galaxies. The primary purpose of this paper is to complete the catalogue of photometric measurements of the GOTOQ sample from SDSS. 



There are several obvious biases. First, these galaxies show up because fibers were pointed at QSO candidates selected from the SDSS QSO candidate follow-up spectroscopy \citep{richards03}. Those that were real QSOs \citep[105783 QSOs as of DR7]{schneider10} show at least 100 galaxies { at $z<0.4$} which are included in the 3\arcsec diameter fibers of SDSS. The question arises whether these galaxies would be missed in a random galaxy survey without the background QSOs. Assuming a homogeneous distribution of galaxies in the sky, pointing the same number of fibers ($\sim109,000$) at random parts of the sky would presumably also give about the same number of galaxies to the same magnitude limit. We find {  galaxies down to a photometric magnitude limit} of $r<23$ and still detect strong spectroscopic emission in one hour of integration on SDSS. Therefore we expect that in a random galaxy survey search the galaxies in our sample would not be passed over based on their emission line strengths, and there is no reason to expect our sample to differ from a random search except for cosmic small number statistics.

The second bias is that the continuum noise from the background QSOs actually limits the detection limit of the survey. The faintest galaxies would not be seen in the brightest QSOs (N10). However, since the number of QSOs detectable to the 19.1 mag photometric limit is roughly a factor of six per magnitude interval, this is a relatively small effect. This is derived from the fact that at low redshifts, any population uniformly distributed in space with a uniform luminosity distribution at all redshifts has six times more objects in an interval of one magnitude than in the previous interval of one magnitude. Simply, the very brightest QSOs are few and therefore the number of GOTOQs missed along the lines of sight to these QSOs is negligible to the total sample. See N10 for a complete discussion. From this bias, $<1\%$ of a true sample is missed, since at 17 mag there are already 1/36 as many QSOs as at 19 mag, and 1/216 as many at 16 mag. 

The galaxies may have significant effect on the colors of the QSOs, causing the QSO to fall outside the QSO-selection color-box. The same exposure time may detect the galaxy in a random fiber position on the sky. Additionally, if the galaxy is of sufficient brightness, it may make it difficult to determine a point source (such as the QSO).

\section{Data Analysis}\label{section-analysis}

The sample herein has been treated in the same way as the samples from Y12 and S13 respectively. For more in depth discussion of these methods, please refer to these papers. In the process of completing the GOTOQ search, all observable properties of the galaxies and the absorption lines and extinctions were determined. 

\subsection{Broad-band flux measurements}\label{section-photometry}

To determine the photometric properties of the QSO and galaxy separate from one another, we have removed the QSO by PSF subtraction in order to avoid flux contamination to either.  { The PSF subtraction was performed with the Image Display Paradigm 3 (IDP3) software. Figure~\ref{fig-thumb-III} shows the SDSS tricolor composite thumbnail of each of our targets with the accompanying pre- and post-PSF subtracted SDSS data. The second panel shows contours of the QSO and galaxy, and the third panel  shows the contours of just the galaxy after PSF subtraction of the QSO. The contours mark arbitrary flux levels in the QSO and galaxy in order to highlight their structure. Residuals at the position of the QSO are insignificant, being $<1\sigma$ in the noise. The full figure can be found in the online version of the text. Figure~\ref{fig-dark-III} shows SDSS tricolor composite thumbnails for those fields where we did not detect any galaxy after PSF subtraction of the QSO.  Table~\ref{tbl-phot-galaxies-III} details the angular offsets of the detected galaxies from the QSOs  along with the impact parameter $b$ in kpc at the redshift derived from the galactic emission lines, assuming the adopted cosmology. }

Reported apparent magnitudes are the asinh magnitudes as prescribed by \cite{lupton99}. Table~\ref{tbl-phot-quasars-III} details the magnitudes calculated for the QSOs in all five filters ($ugriz$). Also presented in this table are $r$-band magnitude PSF model fits to the QSO from SDSS, which are fit and measured pre-deconvolution of the galaxy and QSO. Additionally, the Petrosian radius ($\mathcal{R_{P}}$) for each QSO is provided. The Petrosian radius is a measure of the size of an object, and values larger than about 1.4$\arcsec$ indicate a profile that deviates from a point source. This could be an indicator of a galaxy's presence even if such a galaxy { could not be discerned after QSO deconvolution. The galaxy might be too faint to discern in the photon noise of the wings of the QSO image (too small or too faint) or the object might have very low baryonic mass, but still have emission lines, with little stellar continuum \citep[as in][for instance]{rhode13}.}  Table~\ref{tbl-phot-galaxies-III} details the magnitudes and luminosities for the galaxies in all five filters as well as the galaxy (u-r) color.  {  All magnitudes for QSOs and galaxies have errors of less than 0.05 mag. Where no galaxy is detected, we report magnitude limits of the following: m$_{u}>23.60$, m$_{g}>24.45$, m$_{r}>24.00$, m$_{i}>23.50$, and m$_{z}>23.50$. }

We have also determined the dust and reddening estimates for the QSOs and galaxies. The observer-frame color excess ($\Delta$(g-i)) is calculated by taking the color difference (g-i) for the QSO and comparing it to the median (g-i) for QSOs at the same redshift, taken from \cite{schneider07}. This value allows us to estimate the reddening of the QSO due to dust in the intervening emission galaxy.  Additionally, from this $\Delta$(g-i) value, we have calculated the E(B-V)$_{(g-i)}$ value (the absorber rest-frame color excess) according to the prescription of \cite{york06}. { Several of the QSOs in our sample have multiple absorption line systems. Therefore, reddening values cannot be attributed solely to the galaxy in question in these cases. }

\subsection{Emission line measurements}

We have performed the same measurements as in Papers I and II of this series in order to compare our samples consistently. Therefore, parameters for each of the emission lines found in the spectra were measured with the SPLOT package within IRAF. Figure~\ref{fig-spectra} shows the SDSS spectra for each of our targets. Determined flux values can be found in Table~\ref{tbl-emission-III} measured in 10$^{-17}$ ergs  cm$^{-2}$ s$^{-1}$. Wavelengths are measured in Angstroms. Error values are determined by measuring the standard deviation in a featureless region of the spectra near the emission line and multiplying this value by the square root of the number of pixels occupied by the emission line. These are available for the interested reader. 

Integrated star formation rates (SFR) from H$\alpha$ and [O~II]$\lambda$3727 emission were calculated as prescribed in Kennicutt (1998).  The H$\alpha$/H$\beta$ values allow us to calculate the Balmer decrement extinction correction to SFR, as well as the reddening values for the SMC extinction curve (see \cite{york06}), given by E(B-V)$_{H\alpha/H\beta}$ (see \cite{straka13}). These are given in Table 5 where we have reported the uncorrected SFRs in addition to the SFR values corrected for extinction in the foreground galaxy via the Balmer decrement. We also correct for Milky Way extinction. 

From the measured emission line fluxes, we can determine the emission line metallicity via several techniques, namely the indices R$_{23}$ \citep{pagel79}, N2, and O3N2 \citep{pettini04}. 

\begin{equation}
log R_{23} = log\left( \frac{[O II]\lambda3727 + [O III]\lambda\lambda4959, 5007}{H\beta}\right)
\end{equation}

\begin{equation}
N2 = log \left(\frac{[N II]\lambda6584}{H\alpha}\right)
\end{equation}

\begin{equation}
O3N2 = log \left[\left(\frac{[O III]\lambda5007}{H\beta}\right)\Bigg/\left(\frac{[N II]\lambda6584}{H\alpha}\right)\right]
\end{equation}
From these values we determine the analytic $12+log(O/H)$ values prescribed by \cite{mcgaugh91} and \cite{pettini04}. Table~\ref{tbl-emission-III} lists these values. A full discussion of these measurements can be found in the sections to follow.  {  Limits for our emission measurements are available in the online table.}

\section{Results}\label{section-results}

\subsection{Photometry}

Table~\ref{tbl-phot-quasars-III} lists the PSF r-band magnitude for each QSO, as well as the Petrosian radius $\mathcal{R_{P}}$, which can be used as an indicator of the presence of a galaxy. Petrosian radii greater than $\sim1.4$\arcsec are indicative of the presence of an extended object. Seeing can greatly affect the Petrosian radius, but with few exceptions the majority of our detections have $\mathcal{R_{P}}>1.4$\arcsec. Interestingly, all but two of our fields with undetected galaxies also have $\mathcal{R_{P}}>1.2$\arcsec. This indicates that the galaxies we seek are small and may fall within the PSF of the QSO. High resolution instrumentation in combination with adaptive optics will be required to separate these galaxies from their QSO counterparts. This will be addressed below. 

 Figure~\ref{fig-ur-III} shows a histogram of each of the samples, with a total sample histogram shown in solid black line. The total sample from all three searches (the total sample refers to all 103 GOTOQs together) yields a mean (u-r) value of 1.14 and a median of 1.26. {  \cite{strateva01} present the color separation of galaxy types in SDSS data. In comparison to our color histogram,} Figure 7 of \cite{strateva01} displays the analogous histogram for (u-r) values for their spectroscopic sample. The dividing color value between morphological types according to \cite{strateva01} is (u-r)$=2.22$. Comparison with our own total sample shows that over 80\% of galaxies in our combined sample fall in the range of late-type galaxies. For the total sample, over 90\% fall in the range of late-type galaxies. Two of our targets from this work have (u-r)$>2.22$ (target numbers 28 and 38), indicating they may be good candidates for a green valley study of red spirals or blue ellipticals. One other target from S13 (Q0259+0001) has (u-r)$=2.64$, bringing the total to three in the green valley region.  

The L$^{\ast}_{r}$ values are listed in Table~\ref{tbl-phot-galaxies-III}. Values $<0.2$L$^{\ast}$ are indicative of dwarf galaxies. By this definition, over half of our sample herein could potentially be dwarf galaxies. The connection with dwarf halos and dark galaxies is discussed below. 

{ 
\subsubsection{Stellar mass estimates}\label{section-stellar_mass}

Following the measurement of the five-band magnitudes for our systems, we have estimated the stellar mass of each detected galaxy using the photometric redshift code HyperZ \citep{bolzonella00}. The code allows for fixed spectroscopic redshifts to be input (known from our selection criteria), constraining the spectral energy distribution (SED) model fit to the known redshift. The code then performs a reduced $X^{2}$ analysis to obtain the best fit SED using stellar population synthesis models from \cite{bruzual03}. Recent work by \cite{taylor14} shows that photometric redshift software calculations are not significantly improved by the inclusion of near-infrared data points beyond the initial five-band optical data points. In fact they may decrease the robustness of the calculations. We therefore believe our stellar mass calculations from our five-band photometry are as robust as possible. Our estimates for M$_{\ast}$ are in Table~\ref{tbl-phot-galaxies-III}. 

Figure~\ref{fig-stellar_mass_sfr} shows the relationship we find between our HyperZ stellar masses and the SFR measured within the SDSS spectral fiber (taken as a lower limit as the total galaxy area may not fall entirely within the fiber). All SFR are the H$\alpha$ Balmer decrement corrected SFR where available, except in the $z>0.4$ cases where we use the [O II] SFR. We see good agreement with previous trends in log M$_{\ast}$- log SFR from \cite{elbaz07} and \cite{tasca14}. Overplotted are the relationships at $z\sim0$ and $z\sim1$ from \cite{elbaz07}. We note that our SFR are higher than those cited therein, which is attributed to our selection criteria. Since we have required strong H$\alpha$ or [O III] in our sample, we naturally select stronger star forming regions than in \cite{elbaz07}.  We observe a median trend of

\begin{equation}
log SFR = 2.12 + 0.95 log \frac{M_{\ast}}{10^{11}}
\end{equation}

We use this equation to extrapolate the estimated stellar masses of those galaxies with no imaging detections in our sample. These are depicted in Figure~\ref{fig-stellar_mass_sfr} as circles, also color coded by redshift. These points show a clear increase in log M$_{\ast}$-log SFR with redshift. In order to illustrate the reason these galaxies may not have been detected, we also plot horizontal and vertical lines at the median boundary for each redshift bin. This makes it obvious that for all but the highest redshift non-detections, these galaxies lie below the median log M$_{\ast}$-log SFR for detected galaxies in the same sample. This is particularly evident in the $0.1<z<0.19$ bin where the orange circles are all significantly below the median. In this case, we are confident these galaxies were not detected due to their low mass and SFR.  In the case of $0.3<z<0.39$, the non-detections are scattered about the median, making it more difficult to estimate the cause. As we look to increasing redshift, in the case of the $z>0.5$ bin we see all of our non-detections are significantly above the median for detections in this sample. However, the small sample size of this bin limits the statistical analysis we can make.  
 
}

\subsection{Dust measurements along QSO sight-lines}

The total sample of GOTOQs has a mean $\Delta$(g-i) of 0.11 and a median value of 0.07. From the $\Delta$(g-i) values, we determine E(B-V)$_{(g-i)}$. We have assumed R$_{V}=3.1$. The total sample has a mean E(B-V)$_{(g-i)}$ of 0.06 and a median of 0.05.  We have included in Table~\ref{tbl-phot-quasars-III} the number of additional intervening absorption systems detected in the SDSS spectra, excluding any absorption systems at the same redshift as the emission. Interestingly, the most reddened QSOs in our sample do not correspond to those QSOs with the most absorption systems, indicating the dust for these systems is very probably associated with the galaxy detected in emission. In fact, three of our most reddened QSOs from this work (IDs (E(B-V)): 10 (0.34), 35 (0.36), and 42 (0.37)) have no additional intervening absorption systems detected, while one other highly reddened system (49 (0.95)) has only one additional intervening system. Figure~\ref{fig-ebv-III} shows the histogram of extinction values of the combined sample.  It can be seen that the center of the histogram is offset from zero, indicating a higher reddening value for those QSOs with foreground galaxies on average. \cite{york06} report an average E(B-V)$_{(g-i)}=0.013$ for their Mg II selected sample of 809 QSOs { at higher redshift. As indicated by \cite{york06}, values of E(B-V) this high coming from higher redshift galaxies would produce easily detectable Mg II absorption lines, statistically.} As in {  S13}, we see a spread in the histogram of $\sim$0.2, likely caused by error introduced in the deconvolution of the QSO from the galaxy and the { intrinsic QSO color variations}.  

A KS-test between the total sample (103 data points) and \cite{york06} sample (809 data points) yields a discrepancy of 0.374 and a P-value of 0, indicating these two samples are not from the same parent population. This difference may be due to differences in galaxy type; our galaxies are thought to be primarily dusty late-type chosen by their emission signatures, while the Mg II absorbers from York et al. were chosen by absorption and their emission properties are unknown.  A study of the SFR in the York et al. Mg II sample or a study in Mg II absorption (requiring UV observations) of our GOTOQs would help determine the differences in these two samples. We find our total sample to be much more reddened than the sample of York et al. (2006). If we limit the comparison to only those systems for which we detect Mg II absorption (as in York et al.), we find a discrepancy of 0.338 with a P-value of 8.601\%. However, our sample of Mg II absorbers contained herein is small (only 17 compared to the York et al. sample of 806) and may not give a very accurate result. It is also possible that some of our GOTOQs with galaxies at $z<0.4$ have Mg II absorption, but it is outside  the spectrograph range of SDSS.   

For the combined sample, we find an anti-correlation between impact parameter (b) and E(B-V)$_{(g-i)}$, with a Spearman value of r$_{s}$=-0.27 at the $\alpha=0.03$ significance level, indicating higher dust content in regions closer to the center of the galaxies. Figure~\ref{fig-ebv-b-III}a shows the plot for these variables.
For the total sample, we also find a significant correlation between (u-r) and E(B-V)$_{(g-i)}$, with a Spearman value of r$_{s}=$0.32 at the $\alpha=0.02$ significance level. This is perhaps expected, given that systems with higher dust content will appear more red and those with lower extinction values will be unaffected in their color, remaining blue. Figure~\ref{fig-ebv-ur-III}b plots our data for this relationship.

\subsection{Galactic emission line measurements} 

For the redshift range $z<0.4$, it is possible to detect H$\alpha$, H$\beta$, [O II]$\lambda3727$, [O III]$\lambda\lambda4960, 5008$, [N~II]$\lambda\lambda$6550,6585, and [S~II]$\lambda\lambda$6718, 6733. Table~\ref{tbl-emission-III} reports our emission line detections for each of our systems. Note that we do not apply any correction to these values for reddening or { for} the fact that the entire galaxy may not fall within the spectral fiber. Therefore, these flux values are likely lower limits. In this redshift range, all but one of our systems have detected H$\alpha$ (Q1329+6304, ID 32). This object is near the upper limit of the redshift range and has a noisy spectrum in the region of H$\alpha$. 
For five galaxies at $z<0.4$ (target IDs 26, 27, 40, 41, and 42), we do not detect [O III]$\lambda5008$. However, each of these five have detected H$\alpha$ as a primary indicator. All targets with $z>0.4$ (17 out of 53 targets) have detected [O~II]. Additionally, only targets 3, 10, 28, 32, 43, and 47 lack [O II] detections. 

Table~\ref{tbl-sfr-III} lists the star formation derived using H$\alpha$ and [O II] luminosities. In the 
case of $z<0.4$ we also detected H$\beta$  in 25 cases. In these cases we
derive the dust extinction by comparing the Balmer ratio with the expected value 
of 2.85. The dust corrected star formation rates are also given in Table~\ref{tbl-sfr-III}. 

From H$\alpha$/H$\beta$ we have calculated the E(B-V)$_{H\alpha/H\beta}$. We find a mean value for the total sample of 0.36 with a median of 0.24. In comparing E(B-V)$_{H\alpha/H\beta}$ to E(B-V)$_{(g-i)}$, we find a correlation of r$_{s}$=0.25 with a significance level of $\alpha=0.04$.  As can be seen from Figure~\ref{fig-ebvgi-ebvhahb-III}c, for most of our sample E(B-V)$_{H\alpha/H\beta}$ is higher, suggesting that on average the reddening is higher  around H II regions spread throughout the galaxy (the source of the integrated emission lines) than along the QSO line of sight. We also plot the difference between these two variables, $\Delta$E(B-V), versus impact parameter, which can be found in Figure~\ref{fig-deltaebv-III}d. We find no significant relationship between them.

With few exceptions, the majority of our SFRs are below 5 M$_{\odot}$ yr$^{-1}$. Kobulnicky et al. (2003) present the SFR values for their sample of field galaxies. At redshifts $0.28<z<0.4$, the average SFR is 1.06 M$_{\odot}$ yr$^{-1}$, while the average SFR in the range $0.4<z<0.81$ is 6.01 M$_{\odot}$ yr$^{-1}$. The average SFR for their entire sample of 64 field galaxies in the range $0.28<z<0.81$ is 4.74 M$_{\odot}$ yr$^{-1}$. This value is significantly higher than our average of 2.44 M$_{\odot}$ yr$^{-1}$ over the range of $0<z<0.4$, but includes a much higher redshift range where the SFR is clearly much higher. Their average in the range $0.28<z<0.4$ is comparable to our average, and therefore we suggest that our sample of galaxies is comparable to other samples of field galaxies. Our sample is comparable in SFR to other studies of field galaxies as well \citep{lilly03, mouhcine06}. We require near-infrared spectroscopy on those galaxies at $z>0.4$ to obtain H$\alpha$ SFR values for a more direct comparison between these two samples.

Table~\ref{tbl-emission-III} lists our emission line metallicity values. Emission line metallicities were determined using three indices, including R$_{23}$ \citep{pagel79, kobulnicky99, gharanfoli07}, N2, and O3N2 \citep{pettini04, gharanfoli07}. Where there exists data to determine the ratio of [O~III]$\lambda5008$/[N~II]$\lambda6585$, we can eliminate the degeneracy in the R$_{23}$ index. Where data does not exist for this ratio, we list the lower and upper branch of $12 + log(O/H)$ for R$_{23}$. Following the prescriptions of \cite{mcgaugh91} and \cite{kobulnicky99}, we have calculated the values for $12 + log(O/H)$ for N2 and O3N2 using analytic expressions. We adopted the solar value for $12 + log(O/H)_{\odot}$ to be 8.69 \citep{asplund09}. Our values show that our galaxies are generally below the solar metallicity as measured by R$_{23}$, but generally above solar as measured by O3N2 and N2. We note that our emission line measurements are lower limits, given the fact that the entire galaxy likely does not fall within the SDSS spectral fiber and therefore the fiber does not collect all of the galaxy emission. In comparison with the sample from N10, we see that our values for the lower and upper branches of R$_{23}$ are comparable, as expected.  

{ In the sample from \cite{kobulnicky03}, discussed above ($0.28<z<0.81$),  the average emission line metallicity is found to be }$12+log(O/H)$ of 8.64.  \cite{lilly03} present a sample of 66 galaxies in the redshift range $0.47<z<0.92$. In general, we find our R$_{23}$ upper branch values to be significantly below those found in \cite{lilly03}, but our lower branch values are above theirs. This suggests our sample has a roughly comparable metallicity as theirs, and therefore our galaxy sample is comparable to galaxies at higher redshifts.

Figure~\ref{fig-bpt-III} shows the data points from our sample of 103 GOTOQs (which have detected emission in H$\alpha$, H$\beta$, [O III], and [N II]) on a BPT \citep[Baldwin-Phillips-Terlevich;][]{baldwin81} diagram. The BPT diagram indicates the ionization mechanism of nebular gas in the region observed. The dashed curve represents the division between starburst galaxies and AGN from \cite{kewley01}. The dotted line is the same division, but updated by \cite{kauffmann03}. Additionally, it is accepted that the general definition of Seyfert galaxies is log([O III]/H$\beta$)$>0.48$ and log([N II]/H$\alpha$)$>-0.22$, and the definition for LINERs (low-ionizaton nuclear emission regions) is log([O III]/H$\beta$)$<0.48$ and log([N II]/H$\alpha$)$>-0.22$ \citep{osterbrock89, kewley01, kauffmann03, kewley13}. The horizontal and vertical lines represent these boundaries. 

According to these parameters, all of the galaxies apart from two in our sample with these four emission lines fall within the normal H II, indicating gas is ionized here through normal star formation processes. { The two points falling slightly within the AGN region are within the error bars to be in the H II region as we would expect. Further inspection of their emission lines shows no indication of broadening typical of AGN emission.} Followup spectroscopy will further constrain the positions of these galaxies on the plot and perhaps add new points to the plot through detections of emission lines (particularly [N II]) which are currently undetected.

\subsection{Absorption line measurements}

Measurements of the galaxy absorption features for Ca II K, Ca II H, Na I D2, Na I D1, the Mg II doublet, MgI, and Fe II found in the QSO spectra are reported in Table~\ref{tbl-ew-III} for this work. This table also lists other absorption features detected in the optical in the footnotes, including Mn II, Cr I, Fe II, Ti II, Fe II, and CH, where available. Equivalent width errors are in the range 0.3-0.6\AA. The extension of our search to redshifts up to $z\sim0.8$ allows us to detect these absorption features in some systems in this work, whereas in Y12 and S13 they are not within the range of the spectrograph due to redshift constraints. Additionally, the relative velocity shift between the emission and absorption lines is listed. As in N10, we find the $\Delta$v values to be quite small (all but three $<100$ km s$^{-1}$), which is consistent with the circular velocities of typical galaxies and suggests the absorbing gas is bound to the emission line galaxy. 

Figure~\ref{fig-ca-b-III} shows our data points for the rest equivalent width of Ca II K against impact parameter. We find no significant correlation between these two variables. Similarly, we find no significant correlation between the rest equivalent width of Ca II K and E(B-V)$_{(g-i)}$, displayed in Figure~\ref{fig-ca-ebv-III}. 

The reader should refer to N10 for a full treatment of the Mg II properties of this sample. Fe II $\lambda\lambda2586,2600$ can be seen in the SDSS spectrum at $z>0.47$. Of our 53 targets, 7 are in this redshift range. Of these, 6 have detected Fe II absorption (17, 23, 25, 45, 47, and 51). Similarly, Mg I $\lambda2852$ can only be seen at $z>0.33$; 20 of our 53 targets are in this range, however, only 12 have detected Mg I absorption (2, 3, 6, 8, 22, 23, 24, 29, 45, 47, 50, and 51). 

Those systems with all three detected Mg II, Mg I, and Fe II with equivalent widths $>1$\AA~ fit the criteria set by \cite{rao06} for selecting Damped Lyman-$\alpha$ systems. From our sample, targets number 23, 25, 45, 47, and 51 all  fit this description and { will require followup UV observations to determine if they are indeed} DLA systems. Target number 17 may also be a good candidate, with only Mg I below the threshold. N10 note that it is possible based on the absorption properties, all 17 of the Mg II detected systems could be candidates for DLA systems and 21-cm absorption. { However, based on measurements of strong Mg II and Fe II in \cite{meiring07, meiring08, meiring09a} with Ly$\alpha$ measurement well below the DLA limit, it is unlikely this trend suggested in \cite{rao06} will be supported.} We note, however, that virtually none of the QSOs presented here have sufficient radio fluxes for 21-cm searches with current telescope technology. Additional absorption features found in these systems are noted in the footnotes of Table~\ref{tbl-ew-III}.

In this work, we find 35 of our 53 targets have some form of absorption. Ten of these 35 have no detectable galaxy in imaging after PSF subtraction (IDs 3, 6, 8, 13, 20, 23, 32, 34, 39, and 47). These targets may be good candidates for dark galaxies \citep{matsuoka12, rhode13}. This leaves 18 targets with no detectable absorption, only seven of which have no visible galaxy in imaging (IDs 5, 9, 11, 12, 30, 38, and 44). Comparing the images with detections with the absorption detections, it seems absorption occurs more frequently in the disks of galaxies, while when the impact of the QSO is higher (more considerably in the halo region), absorption is likely to be missed.

{  
\subsubsection{Number density of absorption lines}
We estimate the number density of Ca II, Na I, Mg II, and Mg I by calculating the relative redshift coverage for each of these absorbers in our sample. 

For Ca II, we find a $\Delta Z$ of 8.58, leading to a  d$N$/d$z\sim$0.20. Though we only have 43 Ca II absorbers in our sample, this is higher than dN/dz from \cite{richter11}, who from 18 strong (log N$_{Ca II}>11.65$) Ca II absorbers find a dN/dz of $\sim$0.117 (after bias correction).  We have not selected our systems based on any prior knowledge of their absorption properties and these numbers indicate we have not selected an overabundance of Ca II absorbers in our sample. 

Through similar calculation, we find dN/dz of Mg II for our sample of 18 absorbers to be $\sim$0.50. Again, though limited, this is in agreement with recent studies of Mg II dN/dz such as \cite{evans13} (0.449$\pm$0.003, $0.40<z<$1.14) and \cite{nestor05} (0.60, $z\sim0.3$) for strong Mg II absorbers ($W_{0}^{2796}>1.0$). We cover  a redshift range of $0.36<z<0.79$ with our Mg II absorbers. Mg II absorbers are known to probe a wide array of galactic environments, including tracing outflows, out to impact parameters of 250 kpc \citep[e.g.][]{churchill99, churchill12, lundgren12}. It is therefore no surprise that galaxies known to intervene with background QSOs show strong Mg II absorption at 100\% completion. We refer the reader to \cite{noterdaeme10} for a more complete assessment of the Mg II properties in this sample. 

}

{ 
\subsubsection{Correlations between absorption line equivalent widths}

For our total sample of 103 GOTOQs, we look at the relationship between the strengths of the various absorption lines. In all cases more data points are needed to effectively constrain the relationships. However, we present here the preliminary analysis. We see only two possible correlations: W$^{\lambda3933}_0$ (Ca II K) with W$^{\lambda2852}_0$ (Mg I) and W$^{\lambda2796}_0$ (Mg II) with W$^{\lambda2600}_0$ (Fe II). 

Five systems show both Ca II K and Mg I 2852 (36\% of Mg I have Ca II). This relationship has a Spearman correlation value of 0.70 with an alpha significance of 0.19. However, this is at best a weak correlation and requires many more points for validation. This weak correlation has a relationship of W$^{\lambda2852}_0$ $=2.31$W$^{\lambda3933}_0$ $-0.26$. 

Six of our GOTOQs present with both Mg II and Fe II (100\% of Fe II have Mg II). Even with only six points, we see a preliminary tight correlation between Mg II 2796 and Fe II 2600 equivalent width. This relationship has a Spearman correlation value of 0.77, corresponding to an alpha significance of 0.07. We note that both of these relationships are hindered by our limited low redshift coverage, since Mg II, Mg I, and Fe II are only obtainable in the SDSS spectra at $z>0.39$, $z>0.37$, and $z>0.5$ respectively. This correlation has a relationship of W$_0^{\lambda2600}$~$=0.6$W$^{\lambda2796}_0$ $+0.58$. 

We calculate the relationships between each of the other absorption lines, but find no significant correlation between them. The most significant are the following: 

\begin{itemize}

\item{Fourteen GOTOQ systems have both Ca II K and Na I D2 (33\% of Ca II absorbers have Na I). This results in a Spearman value of 0.25, with alpha significance of 0.46 between the two rest-frame equivalent widths. There is considerable scatter between the two. }


\item{Five systems with both Mg I 2852 and Fe II 2600 (36\% of Mg I absorbers have Fe II). }

\item{Fourteen systems with both Mg II 2796 and Mg I 2852 (78\% of Mg II absorbers have Mg I).  }

\item{Six systems with both Ca II K and Mg II 2796, which is 33\% of Mg II absorbers have Ca II K. This is much higher than the incidence rate found by \cite{sardane14} (3\% of Mg II absorbers have Ca II) and we find no significant correlation between the two in our sample. \cite{sardane14} detect a postive correlation between the strengths of these two absorbers, albeit with a large spread. Our sample, however, is too small for any statistical comparison. We note that our six systems fall well within the trend of \cite{sardane14} Figure 14. The higher incidence rate in our sample could be due to our bias towards very short offsets from the galaxy center, where, in these inner regions, we might expect higher Ca II abundance relative to Mg II and less Ca II ionization.  }

%
%

\end{itemize}

Our sample has too few systems with both Na I D2 and Mg II, Ca II K and Fe II, Na I D2 and Mg I, and Na I D2 and Fe II to make any sensible statistical analysis regarding these absorption relationships.

\subsection{Correlations between emission and absorption strength}

We test to see if there are any correlations between emission and absorption strength in our sample. We compare our strongest emission lines (H$\alpha$, [O~II]$\lambda3727$, and [O~III]$\lambda5008$) with the strongest absorbers within redshift range (Ca II K and Mg II $\lambda2796$). We are not able to determine the relationship between H$\alpha$ and Mg II due to redshift restrictions. We find no correlation between [O~II] and Mg II or [O~III] and Mg II for our limited sample. We find a positive correlation at the 0.03 significance level for H$\alpha$ and Ca II K (Spearman rho of 0.35),  and also and anti correlation between [O~III] and Ca II K at the 0.03 significance level (Spearman rho of -0.30). The best fit to these relationships is as follows: 

\begin{equation}
W^{\lambda3933}_0 = 0.01F_{H\alpha}+0.44
\end{equation}
and
\begin{equation}
W^{\lambda3933}_0 = -0.01F_{[O III]} + 1.14
\end{equation}

Figure~\ref{fig-abs_em} plots our data for these relationships, with the best-fit trend lines overplotted. Strong H$\alpha$ is an  indicator of active star formation, as shown by the SFRs measured for this sample. Likewise, Ca II is a tracer of dusty environments, such as star forming regions, indicating active star formation in dusty regions will have both strong nebular emission and Ca II absorption in related ratios. The opposite trend for [O III] flux may be explained by [O III] not being a direct indicator of SFR, but instead its dependence on n$_{e}$ and T$_{e}$. 

}

\subsection{Dwarf galaxies and dark halos?}

As previously suggested in {  S13}, there is evidence that {  a small number} of our galaxies may be candidates for dark galaxies and a significant fraction of the sample are low-mass dwarf galaxies. 
{  Recent studies by \cite{matsuoka12} show that images as deep as $m_{R}=30$ do not detect the source which is initially detected in HI 21-cm data. They additionally do not detect any nebular emission lines from this dark region, though the optically detected component of this galaxy to the NE does have strong emission detected. Additional data on the 21-cm emission of these continuum sources from the VLA and ALFALFA survey show neutral hydrogen masses as low as 10$^{5}$ M$_{\odot}$. However, in our sample we detect strong nebular emission for all of our targets based on the selection criteria. The findings of \cite{matsuoka12} indicate then that our galaxies with no imaging detection are not candidates for dark galaxies, but instead typical of low-mass dwarf galaxies. 

The SFR for Leo P (Rhode et al. 2013) is log (SFR) $=-4.27$, which is much lower than anything we found in our sample. However, Leo P lies above the L$_{B}$-SFR relation for dark galaxies and therefore, while it is not a candidate for a dark galaxy, may be indicative of the type of objects we looking at in our sample of non-detected galaxies. 

The non-detection of these objects is most probably in large part due to their redshift. The mean and median redshift of our detected galaxies is 0.18 and 0.13 respectively. However, the mean and median redshift of our undetected galaxies is much higher: 0.31 and 0.34 respectively. The higher-$z$ nature of these low-mass objects naturally makes them more difficult to detect in imaging. These objects in particular would not have otherwise been found in traditional galaxy surveys, making them unique to our search technique and sample. If we compare galaxies in our sample with similar SFR, we can see there is a strong dependence on redshift whether these galaxies are detected or not, implying if we moved a detected galaxy to the higher redshift of a non-detected galaxy with similar SFR, we would not detect this galaxy either.  We have also compared the log M$_{\ast}$ values for these non-detected galaxies with our detections at similar redshift. We find that, in addition to having higher than average redshift, they also have lower than average log M$_{\ast}$. }

Currently, the number of observed dwarf galaxies in the universe falls far short of what is predicted by models \citep[e.g.][]{klypin99, moore99}. Additionally, the suppression of star formation in dwarf galaxies is not well understood. These data then show that it is possible to have a gas-rich {  dwarf} galaxy, even at masses as low as 10$^{5}$ M$_{\odot}$, and still detect strong emission and star formation. These qualitative characteristics are comparable to what we have found for our sample of galaxies, strongly suggesting that those fields with undetected galaxies in imaging within our {  sample fall into the category of low-mass dwarf galaxies.   }

It has been suggested in the past that H$_{2}$ density correlates much more closely with star formation than total gas density of a system (Leroy et al. 2008, Bigiel et al. 2008). Kuhlen et al. (2012, 2013) provide model evidence that the suppression of star formation in dwarf and dark galaxies is better controlled by H$_{2}$ gas and results in $\sim80$\% dark halos with M$_{h}<10^{9}$ M$_{\odot}$ and SFR$<1$ M$_{\odot}$ yr$^{-1}$. Thus, it would be interesting to obtain 21-cm and CO data for any of our {  low-mass} dwarf galaxy candidates to see how they compare with these recent findings.

\section{Conclusion}\label{section-conclusion}

{    Here we report 103 galaxies intervening with background QSOs detected in imaging, spectral emission, and absorption. The advantage of this sample is that it is entirely random with respect to the foreground galaxies. The standard procedure of detection of the some $>200,000$ QSO absorption line systems known (York et al. in preparation, Lundgren et al. in preparation) is to note the absorbers in QSO spectra, then look for foreground galaxies. The sample here represents over 40\% of confirmed absorber host galaxies, i.e. those galaxies known to have QSOs probing hypothetical halos of $<200$ kpc in size (even though we find the impact parameters in our survey to all be within $<12$ kpc).  There is no particular reason to believe that QSOs with such close impact parameters would not probe the more extended regions of any halos associated with these galaxies along the line of sight.} The discrepancy in numbers is striking, but possibly just caused by the fact that galaxies as faint as those in our sample (103) have not been reached in the limited searches to date \citep{lundgren12}.

Follow-up of galaxies in this sample should provide a good comparison with high-z samples, to test the current paradigm that halos of galaxies are the cause of QSO absorption lines. Differences of the detected properties of the QSOALS in the two samples may, at any rate, shed light on the detailed structure of galaxy halos.

We have here presented a unique sample of 53 galaxies found intervening between background QSOs in the SDSS DR7 at $z<0.8$ detected primarily via their H$\alpha$ and [O III] galactic emission, bringing our total sample size to 103. We confirmed these galaxies by measuring eight other galactic emission lines where present. The emission and absorption properties of each of these systems has been reported, showing we are studying a sample of late-type galaxies with low SFR. Our targets are shown to be comparable with field galaxies at low redshifts in terms of reddening and SFR. Many of the galaxies in our sample appear to be dwarfs in nature, given their low r-band luminosity. 

We find that the SFR for our targets are, with few exceptions, extremely low. They are, however, in agreement with surveys of field galaxies in the literature. We have also determined emission line metallicity estimates for each of our targets. These values range from sub-solar to super-solar, with the majority being sub-solar in value. These values are also in agreement with studies of field galaxy emission line metallicities. In addition, we have estimated the log M$_{\ast}$ values for our galaxies and extrapolated the relationship to those galaxies with no imaging detections, finding good agreement with the literature. These non-detections are likely low-mass dwarf galaxies. 

We find a significant difference between our total sample and the Mg II selected sample presented in \cite{york06}. Comparing the reddening values of our Mg II absorbers in the total sample to the sample of York et al. (2006), we find an increased likelihood that they could be from the same population, however, the probability is still very small. We do detect an anti-correlation between impact parameter and E(B-V)$_{(g-i)}$ at the $\alpha=0.03$ significance level, as well as a correlation between  E(B-V)$_{(g-i)}$ and (u-r) at the $\alpha=0.2$ significance level. Additionally, we find a correlation between E(B-V)$_{(g-i)}$ and E(B-V)$_{H\alpha/H\beta}$ at the $\alpha=0.04$ significance level. The absorption features show that five (potentially six) of our systems are candidates for DLA systems along the QSO line of sight, with Mg II, Mg I, and Fe II equivalent widths $>1$\AA. 

We estimate the dN/dz values for our Ca II and Mg II absorbers and find compared to other studies of these features we do not have an oversampling, indicating GOTOQs are not preferentially traced by these systems more than expected. Both of these absorption lines are found to trace dusty, star forming regions and a wide range of galactic environments respectively. We have also compared absorption line strengths and found positive correlation between Ca II and Mg I and also between Mg II and Fe II. In comparing emission line strength with absorbers, we find a positive correlation between Ca II and H$\alpha$ and an anti-correlation between Ca II and [O III]. 

We are currently reducing optical integral field spectroscopy of many of our targets in order to determine the total SFR and kinematics of the systems. Additionally, we plan to obtain deeper, higher resolution imaging of those systems with no imaging detections. Also planned are UV spectra of our targets to obtain data on Lyman-$\alpha$ absorption, and Mg II absorption where $z<0.4$. These data will allow a direct comparison with higher-z Mg II absorbers and other absorption systems.

\section*{Acknowledgements}

LAS and VPK would like to acknowledge partial funding from NSF grant AST/0908890 (PI Kulkarni). VPK also acknowledes partial support from NSF grant AST/1108830. LAS and DGY acknowledge support from the John Templeton Foundation through a grant to DGY. We thank the referee for the constructive comments which improved this paper. Funding for the SDSS and SDSS-II has been provided by the Alfred P. Sloan Foundation, the Participating Institutions, the National Science Foundation, the U.S. Department of Energy, the National Aeronautics and Space Administration, the Japanese Monbukagakusho, the Max Planck Society, and the Higher Education Funding Council for England. The SDSS Web Site is http://www.sdss.org/. The SDSS is managed by the Astrophysical Research Consortium (ARC) for the Participating Institutions. The Participating Institutions are The University of Chicago, Fermilab, the Institute for Advanced Study, the Japan Participation Group, The Johns Hopkins University, the Korean Scientist Group, Los Alamos National Laboratory , the Max-Planck-Institute for Astronomy (MPIA), the Max- Planck-Institute for Astrophysics (MPA), New Mexico State University, University of Pittsburgh, University of Portsmouth, Princeton University, the United States Naval Observatory, and the University of Washington.

\bibliographystyle{mn2e}
\bibliography{/Users/straka/Dropbox/bibliography/bibliography}

\clearpage


\begin{figure}
\begin{minipage}{1.0\linewidth}
\begin{center}

\subfigure[1]{
\includegraphics[trim = 0mm 0mm 0mm 0mm, clip, width=0.3\textwidth]{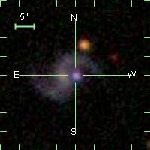}
}
\subfigure{
\includegraphics[trim = 0mm 44mm 0mm 0mm, clip, width=0.6\textwidth]{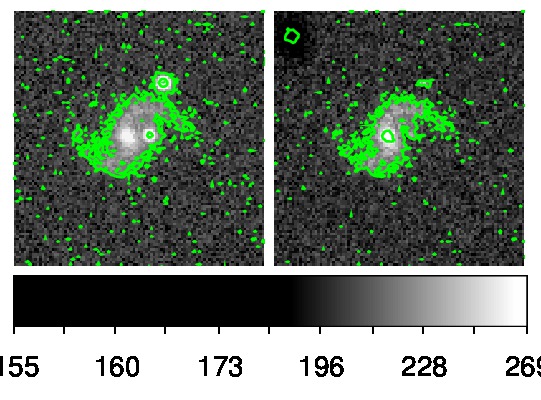}
}
\subfigure[2]{
\includegraphics[trim = 0mm 0mm 0mm 0mm, clip, width=0.3\textwidth]{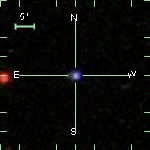}
}
\subfigure{
\includegraphics[trim = 0mm 47mm 0mm 0mm, clip, width=0.6\textwidth]{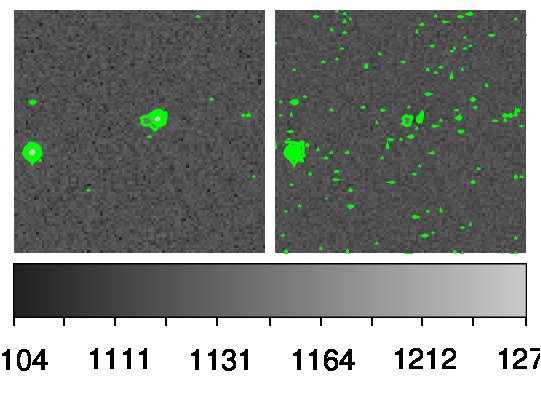}
}
\subfigure[4]{
\includegraphics[trim = 0mm 0mm 0mm 0mm, clip, width=0.3\textwidth]{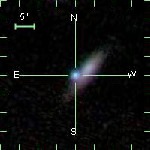}
}
\subfigure{
\includegraphics[trim = 0mm 45mm 0mm 0mm, clip, width=0.6\textwidth]{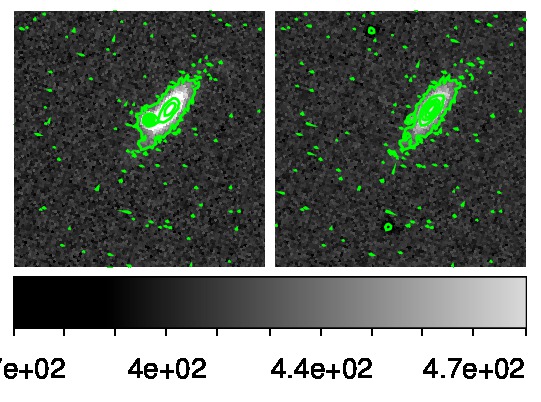}
}

\end{center}
\captcont{SDSS multicolor images of the 36 fields with a QSO intercepting a low-z galaxy detected in imaging, ordered by the increasing RA and increasing numerical index. The scale of the images is indicated in the upper left hand corner. The orientation is north up and east left. On the right, r-band images of the QSO and galaxy before (left) and after (right) PSF subtraction of the QSO. Contours mark arbitrary steps in flux in the image to mark the boundaries of the galaxy and QSO. The full figure may be seen on the web. }\label{fig-thumb-III}
\end{minipage}
\end{figure}

\begin{figure}
\begin{minipage}{1.0\linewidth}
\begin{center}

\subfigure[7]{
\includegraphics[trim = 0mm 0mm 0mm 0mm, clip, width=0.3\textwidth]{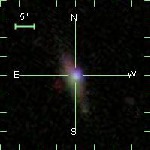}
}
\subfigure{
\includegraphics[trim = 0mm 44mm 0mm 0mm, clip, width=0.6\textwidth]{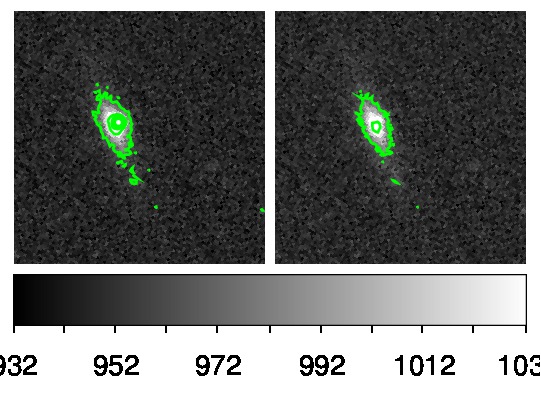}
}
\subfigure[12]{
\includegraphics[trim = 0mm 0mm 0mm 0mm, clip, width=0.3\textwidth]{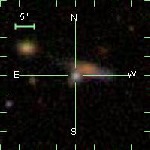}
}
\subfigure{
\includegraphics[trim = 0mm 45mm 0mm 0mm, clip, width=0.6\textwidth]{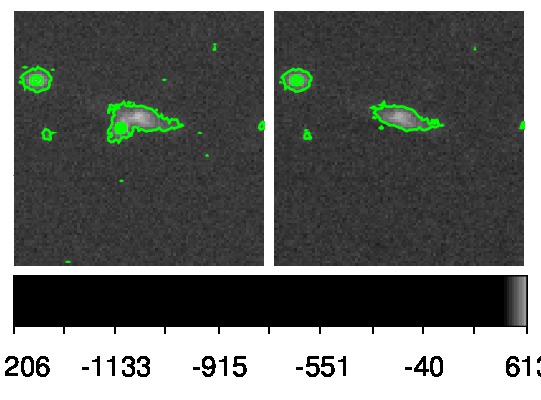}
}
\subfigure[14]{
\includegraphics[trim = 0mm 0mm 0mm 0mm, clip, width=0.3\textwidth]{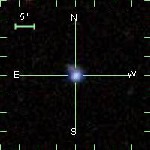}
}
\subfigure{
\includegraphics[trim = 0mm 45mm 0mm 0mm, clip, width=0.6\textwidth]{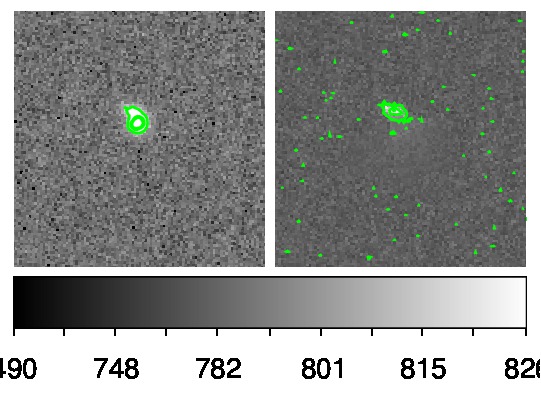}
}

\end{center}
\caption{Continued.}
\end{minipage}
\end{figure}


\begin{figure}
\begin{minipage}{1.0\linewidth}
\begin{center}

\subfigure[3]{
\includegraphics[trim = 0mm 0mm 0mm 0mm, clip, width=0.2\textwidth]{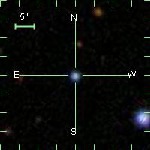}
}
\subfigure[5]{
\includegraphics[trim = 0mm 0mm 0mm 0mm, clip, width=0.2\textwidth]{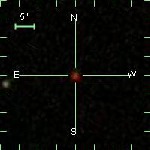}
}
\subfigure[6]{
\includegraphics[trim = 0mm 0mm 0mm 0mm, clip, width=0.2\textwidth]{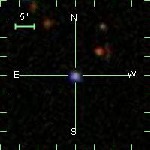}
}
\subfigure[8]{
\includegraphics[trim = 0mm 0mm 0mm 0mm, clip, width=0.2\textwidth]{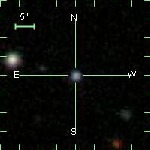}
}
\subfigure[9]{
\includegraphics[trim = 0mm 0mm 0mm 0mm, clip, width=0.2\textwidth]{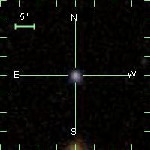}
}
\subfigure[11]{
\includegraphics[trim = 0mm 0mm 0mm 0mm, clip, width=0.2\textwidth]{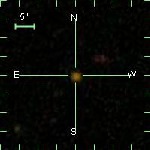}
}
\subfigure[12]{
\includegraphics[trim = 0mm 0mm 0mm 0mm, clip, width=0.2\textwidth]{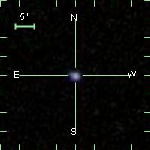}
}
\subfigure[13]{
\includegraphics[trim = 0mm 0mm 0mm 0mm, clip, width=0.2\textwidth]{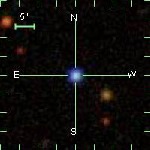}
}
\subfigure[20]{
\includegraphics[trim = 0mm 0mm 0mm 0mm, clip, width=0.2\textwidth]{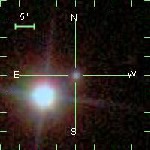}
}
\subfigure[23]{
\includegraphics[trim = 0mm 0mm 0mm 0mm, clip, width=0.2\textwidth]{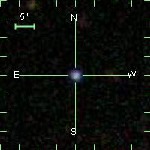}
}
\subfigure[30]{
\includegraphics[trim = 0mm 0mm 0mm 0mm, clip, width=0.2\textwidth]{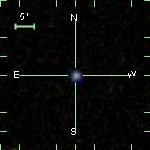}
}
\subfigure[32]{
\includegraphics[trim = 0mm 0mm 0mm 0mm, clip, width=0.2\textwidth]{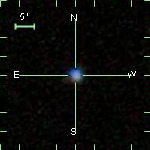}
}
\subfigure[38]{
\includegraphics[trim = 0mm 0mm 0mm 0mm, clip, width=0.2\textwidth]{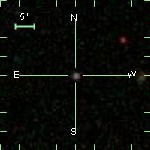}
}
\subfigure[39]{
\includegraphics[trim = 0mm 0mm 0mm 0mm, clip, width=0.2\textwidth]{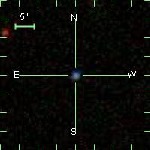}
}
\subfigure[44]{
\includegraphics[trim = 0mm 0mm 0mm 0mm, clip, width=0.2\textwidth]{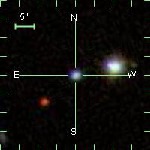}
}
\subfigure[45]{
\includegraphics[trim = 0mm 0mm 0mm 0mm, clip, width=0.2\textwidth]{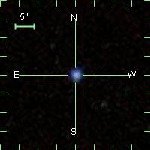}
}
\subfigure[47]{
\includegraphics[trim = 0mm 0mm 0mm 0mm, clip, width=0.2\textwidth]{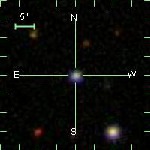}
}

\end{center}
\caption{GOTOQs without visually confirmed galaxies. These are candidates for dwarf and dark galaxies. \label{fig-dark-III}}
\end{minipage}
\end{figure}


\begin{figure}
\begin{minipage}{1.0\linewidth}
\begin{center}

\includegraphics[trim = 0mm 0mm 0mm 0mm, clip, width=0.65\linewidth]{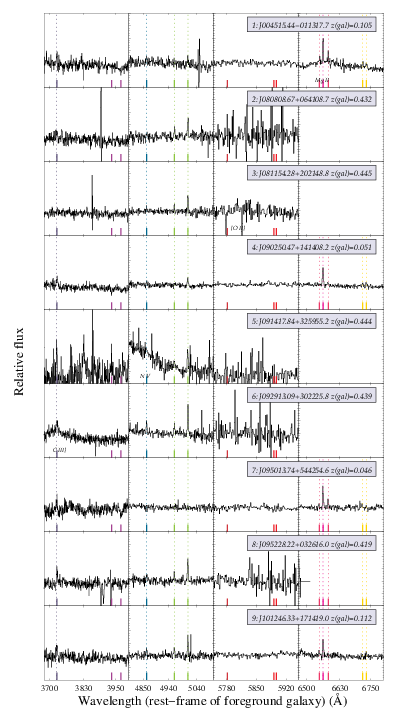}

\end{center}
\caption{SDSS spectra of our targets. The colored tick marks at the bottom of each spectra mark the expected positions of each emission and absorption feature. In order, these are [O II]; Ca II; H$\beta$; [O III]a,b; the 5780\AA ~dusty interstellar band (DIB) feature; Na I; [N II]a; H$\alpha$; [N II]b; and [S II]a,b. Broad emission lines labeled at the base of the spectra (such as Mg II in the first spectra) indicate QSO emission. Other absorption features not marked are from systems at other redshifts. The remaining spectra figures may be seen in the web version of this article. \label{fig-spectra}  }
\end{minipage}
\end{figure}


\begin{figure}
\begin{minipage}{1.0\linewidth}
\begin{center}
\subfigure[]{\label{fig-ebv-III}
\includegraphics[trim = 0mm 0mm 0mm 0mm, clip, width=0.75\linewidth]{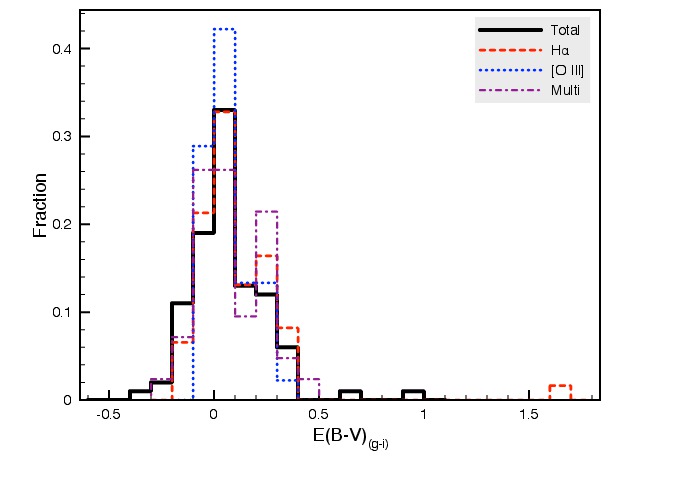}
}
\subfigure[]{\label{fig-ur-III}
\includegraphics[trim = 0mm 0mm 0mm 0mm, clip, width=0.75\linewidth]{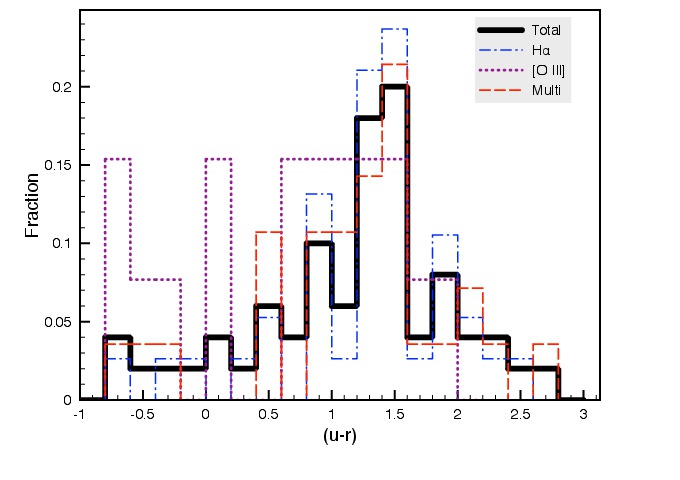}
}
\end{center}
\caption{a) Histogram of E(B-V)$_{(g-i)}$.  b) Histograms of galaxy color for each of our individual samples and the combined sample. The red dashed-and-dotted line is for the H$\alpha$ selected sample. The purple dashed line is for the multi-emission selected sample. The blue dotted line indicates the histogram for the sample new to this paper. The black solid line is for the combination of all three samples. }
\end{minipage}
\end{figure}


\begin{figure}
\begin{minipage}{1.0\linewidth}
\begin{center}
\subfigure[]{
\includegraphics[trim = 0mm 0mm 0mm 0mm, clip, width=0.45\linewidth]{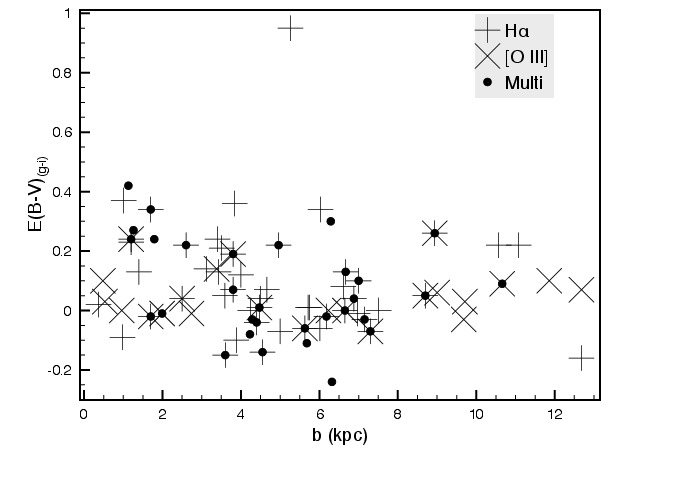}
\label{fig-ebv-b-III}
}
\subfigure[]{
\includegraphics[trim = 0mm 0mm 0mm 0mm, clip, width=0.45\linewidth]{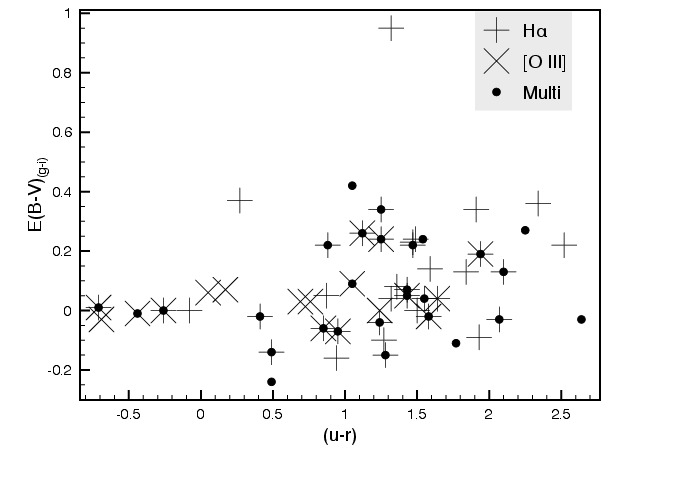}
\label{fig-ebv-ur-III}
}
\subfigure[]{
\includegraphics[trim = 0mm 0mm 0mm 0mm, clip, width=0.45\linewidth]{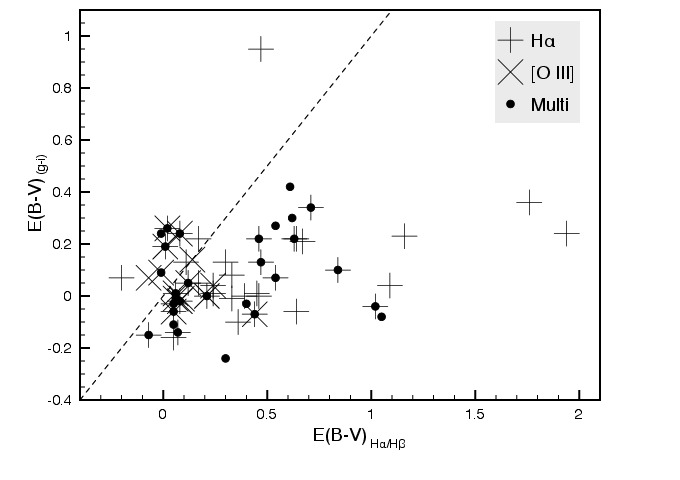}
\label{fig-ebvgi-ebvhahb-III}
}
\subfigure[]{
\includegraphics[trim = 0mm 0mm 0mm 0mm, clip, width=0.45\linewidth]{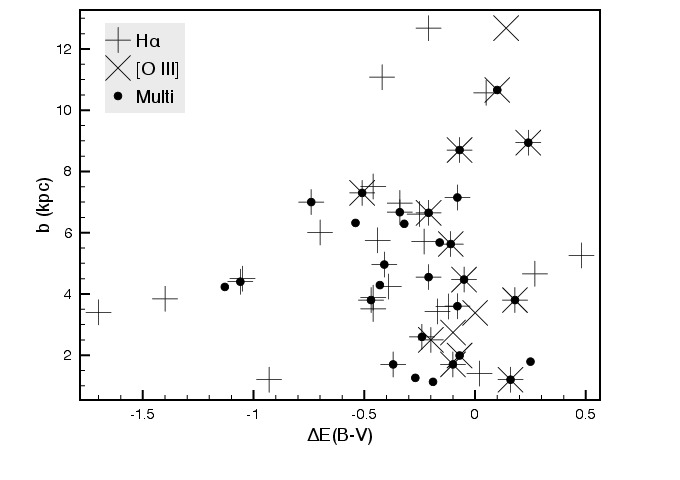}
\label{fig-deltaebv-III}
}
\end{center}
\caption{a) E(B-V)$_{(g-i)}$ against impact parameter. b) Reddening against galaxy color.  c) Both reddening values plotted against one another. d) The difference in our reddening values against impact parameter. }
\end{minipage}
\end{figure}


\begin{figure}
\begin{minipage}{1.0\linewidth}
\begin{center}
\includegraphics[trim = 0mm 0mm 0mm 0mm, clip, width=1.0\textwidth]{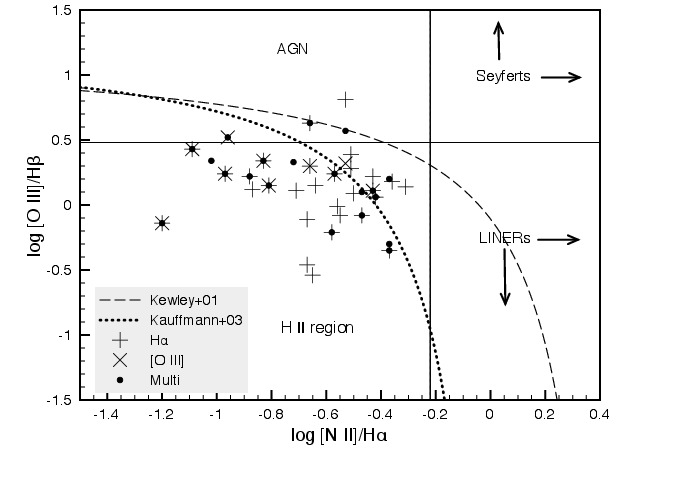}
\end{center}
\caption{BPT diagram (Baldwin et al. 1981) of all three samples in this series. The figure shows that our systems are nearly all normal HII regions, with a few that may be candidates for normal AGN. The dotted line denotes the model of Kauffmann et al. (2003), while the dashed line is the model of Kewley et al. (2001) for the division between starburst galaxies and AGN. The crosshairs mark the division between Seyferts and LINERs.   \label{fig-bpt-III}}
\end{minipage}
\end{figure}


\begin{figure}
\begin{minipage}{1.0\linewidth}
\begin{center}
\includegraphics[trim = 0mm 0mm 0mm 0mm, clip, width=1.0\linewidth]{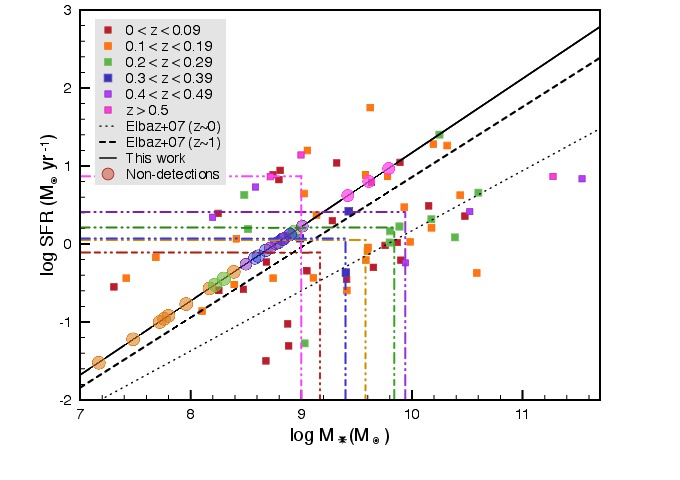}
\end{center}
\caption{Stellar mass of our image-detected GOTOQs as a function of their H$\alpha$ star formation rate are shown as squares. We bin them in 0.1 redshift increments, shown by the different colors. The circles show our extrapolated stellar masses for those galaxies with no imaging detections. We see no real trend in this small sample with redshift for the galaxies with detections, but it is clear that those galaxies with no detections have increasing estimated masses with increasing redshift. We have plotted the observed trend of stellar mass with SFR from \protect\cite{elbaz07}. The $z\sim$1 trend line has a better fit to our median trend line. \label{fig-stellar_mass_sfr}}
\end{minipage}
\end{figure}


\begin{figure}
\begin{minipage}{1.0\linewidth}
\begin{center}
\subfigure[]{\label{fig-ca-b-III}
\includegraphics[trim = 0mm 0mm 0mm 0mm, clip, width=0.8\linewidth]{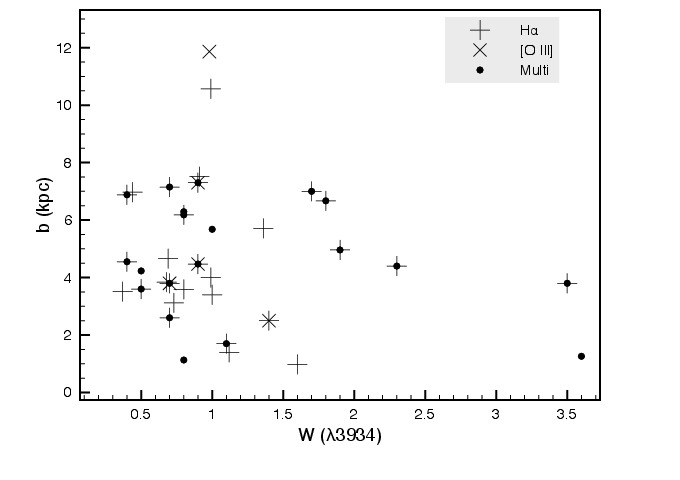}
}
\subfigure[]{\label{fig-ca-ebv-III}
\includegraphics[trim = 0mm 0mm 0mm 0mm, clip, width=0.8\linewidth]{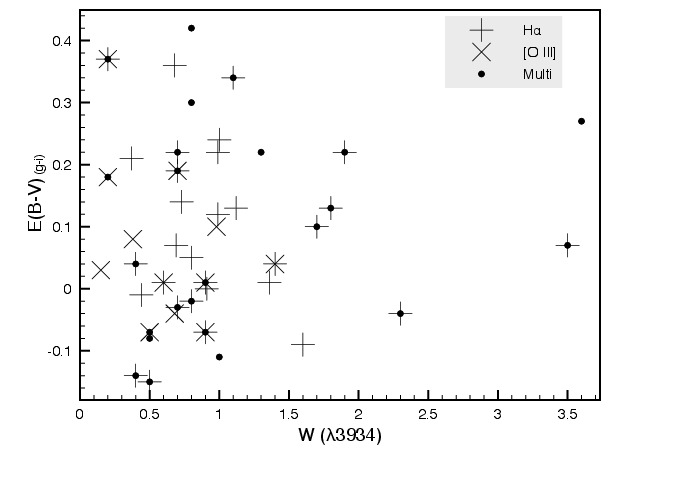}
}
\end{center}
\caption{Ca II K rest equivalent width vs. impact parameter and Ca II K rest equivalent width vs. E(B-V)$_{(g-i)}$. }
\end{minipage}
\end{figure}


\begin{figure}
\begin{minipage}{1.0\linewidth}
\begin{center}
\subfigure[]{
\includegraphics[trim = 0mm 0mm 0mm 0mm, clip, width=0.8\linewidth]{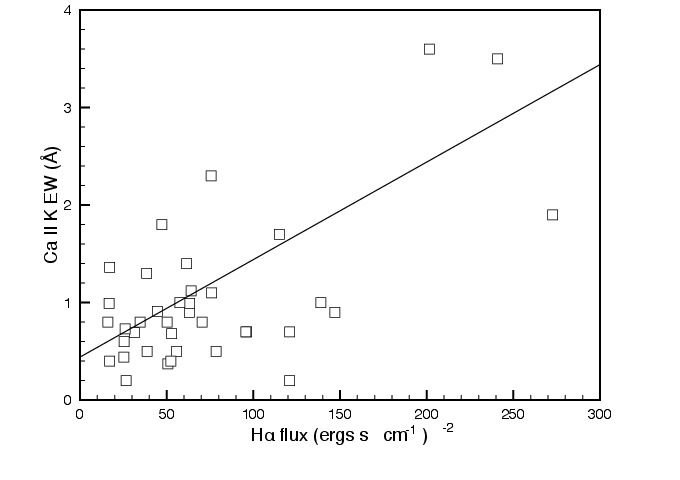}
}
\subfigure[]{
\includegraphics[trim = 0mm 0mm 0mm 0mm, clip, width=0.8\linewidth]{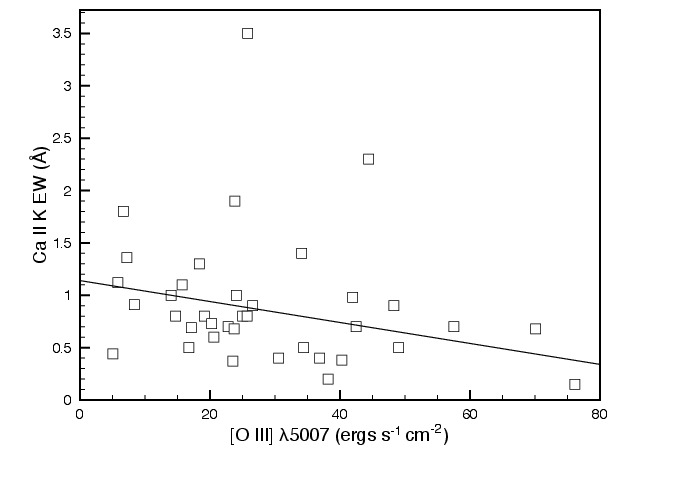}
}
\end{center}
\caption{a). Ca II K rest equivalent width vs. H$\alpha$ flux. We observe a positive correlation between these two parameters. b). Ca II K rest equivalent width vs. [O III]$\lambda$5008. We also observe an anti-correlation between these two parameters. The correlations are both return a significance of 0.03. The plotted lines show our best-fit to each of the sets of data. Please see the text for more details. \label{fig-abs_em} }
\end{minipage}
\end{figure}

\clearpage 

\setlength\tabcolsep{5pt}
\begin{table}
\centering
\begin{minipage}{200mm}
\caption[Catalog of GOTOQs]{Catalog of GOTOQs.\footnotemark}\label{tbl-total-sample}
\begin{tabular}{lllllllllll}

\hline\hline

RA & Dec & Plate & MJD & Fiber & $z_{qso}$ & $z_{gal}$ & Detection\footnotemark & Photo\footnotemark & Search Return\footnotemark \\ \\ \hline\hline 

\hline\hline

00:13:42.44 & -00:24:12.6 & 389 & 51795 & 274 & 1.64 & 0.1557 & York+12 & York+12 & Y \\
00:16:37.07 & +13:56:53.5 & 752 & 52251 & 43 & 2.57 & 0.0912 & York+12 & York+12 & Y,S,N \\
00:45:15.45 & -01:13:17.6 & 393 & 51794 & 122 & 1.59 & 0.1055 & H$\alpha$ & This work. & N \\
00:59:40.67 & -00:09:46.2 & 1083 & 52520 & 105 & 2.39 & 0.3058 & Straka+13a & Straka+13a & S \\
02:37:04.61 & +27:52:39.7 & 2444 & 54082 & 194 & 0.97 & 0.3368 & Straka+13a & Straka+13a & S \\
02:39:14.40 & -07:05:57.1 & 455 & 51909 & 562 & 0.71 & 0.3423 & Noterdaeme+10 & York+12 & Y,N \\
02:43:28.86 & +00:38:31.2 & 807 & 52295 & 341 & 2.75 & 0.0279 & Straka+13a & Straka+13a & S \\
02:59:42.42 & +00:01:38.1 & 802 & 52289 & 244 & 2.20 & 0.0430 & Straka+13a & Straka+13a & S \\
03:12:55.98 & -00:14:00.0 & 1179 & 52637 & 99 & 1.03 & 0.1147 & Straka+13a & Straka+13a & S \\
03:32:28.25 & -00:14:34.7 & 1063 & 52591 & 236 & 1.74 & 0.0030 & Straka+13a & Straka+13a & S \\

\hline

\end{tabular}

\footnotetext{$^1$The full table is available online. }
\footnotetext{$^{2}$First published detection. H$\alpha$ indicates it is not yet published, but is first presented in this paper. \\ York+12 detected in H$\alpha$, Straka+13a detected in 9 multiple galactic emission lines including H$\alpha$, and \\ Noterdaeme+10 detected in [O III].} 
\footnotetext{$^3$First published photometric measurement.} 
\footnotetext{$^4$Notes which automated searches returned these targets. York et al. (2012; Y), Straka et al. (2013, S),\\ and Noterdaeme et al. (N).} 
\footnotetext{$^5$Borthakur et al. (2010).}

\end{minipage}

\end{table}


\begin{table}\footnotesize
\centering
\begin{minipage}{200mm}
\caption{Photometric data for QSOs\label{tbl-phot-quasars-III}}
\begin{tabular}{llccc|ccccccccc}

\hline\hline
& & \multicolumn{2}{c}{SDSS quantities} & \multicolumn{5}{c}{Deconvolved apparent magnitudes} & \multicolumn{2}{c}{Reddening Estimates}  & QSOALS$^3$\\
Index &  ID & m$_{psf(r)}^{1}$ & Petrosian$^{2}$  & m$_{u}$ & m$_{g}$ & m$_{r}$ & m$_{i}$ & m$_{z}$  & $\Delta$(g-i) & E(B-V)$_{(g-i)}$ \\

\hline

1 & Q0045-0113  & 19.33 & 8.708 & 19.91 & 19.44 & 19.41  & 19.01 & 18.94 & 0.13 & 0.08   & 1\\
2  & Q0808+0641 & 19.26 & 0.958 & 19.45 & 19.34 & 19.18  & 19.04 & 18.89  & 0.06 & 0.03 & 4\\
3  & Q0811+2021 & 19.10 & 1.133 & 19.52 & 19.49 & 19.10  & 19.09 & 19.04  & 0.14 & 0.06 & 0 \\
4  & Q0902+1414 & 18.43 & 1.354 & 18.95 & 18.59 & 18.36  & 18.38 & 18.29  & 0.07 & 0.05 & 0\\
5  & Q0914+3259 & 21.41 & 1.439 & -- & -- & 21.72  & 20.04 & 19.08  & -- & -- & 0\\
6  & Q0929+3022 & 19.50 & 1.266 & 19.73 & 19.55 & 19.48  & 19.23 & 19.15  & 0.03 & 0.01 & 0 \\
7   & Q0950+5442 & 18.33 & 3.923 & 18.94 & 18.58  & 18.63  & 18.21 & 18.38 & -0.14 & -0.09  & 3\\
8  & Q0952+0326  & 19.31 & 1.507 & 21.27 & 19.76 & 19.19  & 19.21 & 19.16  & 0.29 & 0.13 & 0\\
9  & Q1012+1714 & 19.09 & 1.548 & 20.07 & 19.52 & 19.09  & 18.97 & 19.10 & 0.35 & 0.21 & 1\\
10  & Q1013-0009 & 19.12 & 2.629 & 20.40 & 19.61 & 18.73  & 18.50 & 18.54 & 0.57 & 0.34 & 0 \\

\hline

\end{tabular}

\footnotemark[1]{PSF fit.}
\footnotemark[2]{Petrosian radius measured by SDSS before deconvolution in arcseconds.}
\footnotemark[3]{The number of intervening absorption systems found in the optical QSO spectra, excluding the system associated with the  galaxy emission. }

\end{minipage}
\end{table}


\begin{table}\footnotesize
\centering
\begin{minipage}{200mm}
\caption{Photometric data for deconvolved galaxies\label{tbl-phot-galaxies-III}}
\begin{tabular}{cccccccccccccccccc}

\hline\hline

Index & $\theta^{1}$ (\arcsec) & b (kpc) & m$_{u}$ & log L/L$_{\sun u}$ & m$_{g}$  & m$_{r}$ & log  L/L$_{\sun r}$ & L$^{\ast}$/L$_{\sun r}$ & m$_{i}$ & log L/L$_{\sun i}$ & m$_{z}$   & (u-r)  & log M$_{\ast}$\\

\hline

1 & 3.43 & 6.61 & 18.90 & 10.36 & 17.94  & 17.54 & 10.24  &  1.07 &  17.29 & 10.28 & 16.96  & 1.36 & 9.62$^{+0.78}_{-0.02}$   \\
2 & 0.09 & 0.53 & 23.33 & 8.59 & 21.94 & 22.64 & 8.20  & 0.01 & 21.45 & 8.62  & 21.02   & 0.69 & 9.94$^{+0.01}_{-0.06}$ \\

3 & -- & -- & -- & -- & -- & -- & -- & -- & -- & -- & -- & --  &  --\\
4  & 3.60 & 3.59 & 18.28 & 10.61 & 17.86 & 17.41 & 10.29  & 1.21  & 17.03 & 10.38  & 16.93  & 0.87 & 9.65$^{+0.03}_{-0.82}$ \\
5 & -- & -- & -- & -- & -- & -- & -- & -- & -- & -- & -- & -- & 8.72$^2$  \\

6 & -- & -- & -- & -- & -- & -- & -- & -- & -- & -- & -- & --  & 8.77$^2$\\
7 & 1.08 & 0.98 & 20.07 & 9.14  & 19.01  & 18.14 & 9.24 & 0.11 & 18.14 & 9.24 & 17.64  & 1.93 & 8.80$^{+0.15}_{0.06}$  \\

8 & -- & -- & -- & -- & -- & -- & -- & -- & -- & -- & -- & --  & 9.0$^2$\\

9 & -- & -- & -- & -- & -- & -- & -- & -- & -- & -- & -- & --  & 7.48$^2$\\
10 & 3.30 & 6.03 & 20.08 & 9.89 & 18.70   & 18.17 & 9.99 & 0.60 & 17.62 & 10.15 & 17.17  & 1.91 & 10.48$^{+0.02}_{-0.13}$ \\

\hline

\end{tabular}

\footnotemark[1]{Dash means that no offset could be determined. Errors on magnitudes are within 0.05 mag.}\\
\footnotemark[2]{We use the relationship between log M$_{\ast}$ and log SFR for our detected systems to extrapolate the log M$_{\ast}$ values for those systems with no imaging detection. Errors for these measurements are within 1 dex. }

\end{minipage}
\end{table}


\begin{table}
\centering
\begin{minipage}{200mm}
\caption{Star formation rates \label{tbl-sfr-III}}
\begin{tabular}{lcccccccccc}

\hline\hline

Index  & E(B-V)$_{H\alpha/H\beta}$ & SFR$_{H\alpha}^{1}$ & SFR$_{[O II]}^{1}$ &SFR$_{H\alpha}^{2}$ &  SFR$_{[O II]}^{2}$   \\
 & & (M$_{\sun}$ yr$^{-1}$)& (M$_{\sun}$ yr$^{-1}$)& (M$_{\sun}$ yr$^{-1}$)& (M$_{\sun}$ yr$^{-1}$) \\

\hline

1 & 0.33 & 0.08 & 0.17  & 0.18  & 0.67 \\
2& -- & -- & 0.58 & -- & --  \\
3 & -- & -- & -- & -- & --   \\
4  & 0.17  & 0.02 & 0.04  & 0.03 & 0.07   \\
5  & -- & -- & 0.89  & -- & --   \\
6  & -- & -- & 1.01 & -- & --   \\
7 & -- & -- & -- & -- & -- \\
8 & -- & -- & 1.69  & -- & --   \\
9  & -0.10 & 0.06 & 0.24  & 0.04 & 0.16 \\
10  & -- & 0.07 & -- & -- & -- \\

\hline

\end{tabular}

\footnotemark[1]{SFR uncorrected for extinction.}
\footnotemark[2]{SFR corrected for extinction using H$\alpha$/H$\beta$.}

\end{minipage}
\end{table}


\begin{table}\footnotesize
\centering
\begin{minipage}{200mm}
\caption[Emission line strengths for galaxies in front of QSOs and emission line metallicities.]{Emission line strengths for galaxies in front of QSOs and emission line metallicities.}\label{tbl-emission-III}
\begin{tabular}{lccccccccccccccccccc}

\hline\hline

ID  & $f_{H\alpha}^{1}$  & $f_{H\beta}$  & $f_{[O II]}$  & $f_{[O III]a}$    & $f_{[O III]b}$   & $f_{[N II]a}$   & $f_{[N II]b}$   & $f_{[S II]a}$  & $f_{[S II]b}$ & \multicolumn{4}{c }{$12 + log(O/H)$}  \\
& & & & & & & & & &  R$_{23 l}$  & R$_{23 u}$ & O3N2 & N2 \\

\hline

1 	& 37.73	          & 9.74		&43.49			&-- $^{2}$		& 4.02		&15.68	&6.51		&8.56		& 14.79  & -- & -- & 8.86  & 8.91\\
2 &	--			&13.97		&6.04			&14.82		&20.54		&--		&--			&--			&-- & 7.27 & 8.91 & -- & --\\
3 &	--			 &	5.28		&	--			&8.67		&20.65		&--		&--			&--			& -- & -- & -- & -- &-- \\
4 	&50.32		&15.01		&43.02		         &--			&25.03		&--		&18.83		&--			& -- & -- & -- & 8.67  & 8.91 \\		
5 	 &--			&--			&8.75			&3.33		&15.99		&--		&--			&--			&-- & -- & -- & -- & --\\
6 	&--			&10.69		&10.16			&11.39		&29.20		&--		&--			&--			&-- & 7.63 & 8.77 &--  &--\\
7   & -- & -- & -- & -- & -- & -- & --&  --&  -- & --  & --  & -- & --  \\
8 	 &--			 &	8.63		&19.01			&6.53		&22.23		&--		&--			&--			&-- & 7.89 & 8.77 & -- &--\\
9 	&21.68		&8.24		&53.96			&20.28		&22.74		&4.94	&--		      	&--			&6.16 & 8.59 & 8.14 & -- &--  \\
10 	 &38.10		&--			&--				&5.91		&15.68		&7.96	&11.44		&20.39		&24.52 & -- &-- & -- & 8.92 \\

\hline

\end{tabular}

\footnotemark[1]{Flux measured in 10$^{-17}$ ergs s$^{-1}$ cm$^{-2}$.}\\
\footnotemark[2]{Indicates no emission line detected or emission line is redshifted out of range of the spectrograph.}

\end{minipage}
\end{table}


\begin{table}\footnotesize
\centering
\begin{minipage}{170mm}
\caption{Rest-frame equivalent widths (W) for absorption interstellar lines in foreground galaxies.}\label{tbl-ew-III}
\begin{tabular}{ccccccccccccccc}

\hline\hline

Index & z$_{abs}$ & $\Delta$v& Ca II K  & Ca II H  & Na I D2  & Na I D1 & Mg II &  Mg II & Mg I & Fe II & Fe II   \\
& & km s$^{-1}$& W$_{\lambda3933}$  & W$_{\lambda3969}$  & W$_{\lambda5890}$  & W$_{\lambda5896}$  & W$_{\lambda2796}$  & W$_{\lambda2804}$  & W$_{\lambda2852}$ & W$_{\lambda2587}$ & W$_{\lambda2600}$   \\

\hline

1 & 0.1059 & 115.2  & $<0.39$  &$<0.63$  & 0.36  & 0.33 & --  & --& --  & --& --   \\
2 & 0.4327 & 72.9 &$<0.63$  &$<0.65$  &$<1.53$  & $<1.53$  & 2.42  & 1.94 & 0.80  & -- & --    \\
3 & 0.4448 & 32.2 &$<0.42$  &$<0.34$   &$<1.05$  & $<1.05$  & 0.98  & 1.05  & 0.30 & -- & -- \\
4 & 0.0504 & -18.4 & 0.80  & 0.39 & $<0.35$  & $<0.35$ & --  & -- & -- & -- & -- \\
5 & -- & -- &$<4.23$  &$<3.58$ &$<3.38$  & $<3.38$ & -- &--  & -- & -- & -- \\
6$^{1}$ &  0.4391 & -58.1 &$<0.76$ &$<0.67$ & 1.94$^6$  & 1.94$^6$  & 1.99  & 1.49 & 1.47 & -- & --   \\
7 & 0.0461 & 0.00 & 1.6  & 1.8  &--  & -- & -- & -- & -- & -- & -- \\
8 &  -- & -- &$<0.56$  &$<0.45$ &$<1.54$  & $<1.54$  & 2.00  & 0.60  & 2.50 & -- & -- \\
9 &  -- & -- &$<0.50$  &$<0.60$  &$<0.40$  & $<0.40$ & --  & -- & -- & -- & -- \\
10 &  -- & -- &$<0.78$ &$<0.78$  &$<0.55$  &$<0.55$ & --  & -- & -- & -- & -- \\

\hline

\end{tabular}

\footnotemark[1]{Also:  CH, W$_{4234}=1.19$ }
\footnotemark[2]{Also:  Mn II, W$_{\lambda2594.50}=0.77$; Cr I, W$_{\lambda3579.70}=0.68$; Cr I, W$_{\lambda3594.50}=0.63$}
\footnotemark[3]{Also: Ti II, W$_{\lambda3242.90}=1.03$}
\footnotemark[4]{Also:  Ti II, W$_{\lambda3242.90}=0.47$}
\footnotemark[5]{Also:  Mn II, W$_{\lambda2606.50}=$0.32; Ti II, W$_{\lambda3384.70}=$0.58; Fe II, W$_{\lambda2344.20}=$1.81; Fe II, W$_{\lambda2374.50}=$1.55; Fe II, W$_{\lambda2382.80}=$2.09 }
\footnotemark[6]{These absorption features are blended and measured together.}

\end{minipage}
\end{table}

\bsp

\label{lastpage}

\end{document}